\documentclass[a4paper,pre,reqno,superscriptaddress,onecolumn,showpacs,footinbib,10pt]{revtex4-2}
\usepackage{geometry}
\usepackage[utf8]{inputenc} 
\usepackage[T1]{fontenc}    
\usepackage{hyperref}       
\usepackage{url}            
\usepackage{booktabs}       
\usepackage{amsfonts}       
\usepackage{lipsum}
\usepackage{graphicx}
\usepackage{amsmath}
\usepackage{amssymb}
\usepackage{xcolor}
\usepackage{graphicx}
\usepackage{bbold}
\usepackage{setspace} 
\usepackage{todonotes} 
\graphicspath{ {./images/} }
\usepackage{amsmath,amssymb,amsfonts,upref,endnotes}
\usepackage{amsthm}
\usepackage{todonotes}
\usepackage{changes}
\usepackage{soul}
\RequirePackage{graphicx}
\usepackage[figuresright]{rotating}
\usepackage{xcolor}
\usepackage{bbold}

\def\bea{\begin{eqnarray}}
\def\eea{\end{eqnarray}}
\def\la{\langle}
\def\ra{\rangle}

\newcommand{\eref}[1]{Eq.~\eqref{#1}}

\begin{document}

\title{Harmonic chain driven by active Rubin bath: transport properties and steady-state correlations}

\author{Ritwick Sarkar}
\email{ritwick.sarkar@bose.res.in}
\affiliation{S. N. Bose National Centre for Basic Sciences, Kolkata 700106, India}

\author{Ion Santra}
\email{ion.santra@theorie.physik.uni-goettingen.de}
\affiliation{Institut für Theoretische Physik, Georg-August-Universität Göttingen, Goettingen 37077,  Germnay}

\author{Urna Basu}
\email{urna@bose.res.in}
\affiliation{S. N. Bose National Centre for Basic Sciences, Kolkata 700106, India}


\begin{abstract}
Characterizing the properties of an extended system driven by active reservoirs is a question of increasing importance. Here we address this question in two steps. We start by investigating the dynamics of a probe particle connected to an `active Rubin bath'---a linear chain of overdamped run-and-tumble particles. We derive exact analytical expressions for the effective noise and dissipation kernels, acting on the probe, and show that the active nature of the bath leads to a modified fluctuation-dissipation relation. In the next step, we study the properties of an activity-driven system, modeled by a chain of harmonic oscillators connected to two such active reservoirs at the two ends. We show that the system reaches a nonequilibrium stationary state (NESS), remarkably different from that generated due to a thermal gradient. We characterize this NESS by computing the kinetic temperature profile, spatial and temporal velocity correlations of the oscillators, and the average energy current flowing through the system. It turns out that, the activity drive leads to the emergence of two characteristic length scales, proportional to the activities of the reservoirs. Strong signatures of activity are also manifest in the anomalous short-time decay of the velocity autocorrelations. Finally, we find that the energy current shows a non-monotonic dependence on the activity drive and reversal in direction, corroborating previous findings. 
\end{abstract}



\maketitle

\tableofcontents

\section{Introduction}
Nonequilibrium reservoirs, defying the fluctuation-dissipation relation (FDR)~\cite{Kubo}, show more complex behavior compared to their equilibrium counterparts~\cite{maes2013,maes2014,maes2015,vandebroek,Steffenoni2016}. Recent years have seen an increasing effort to model and characterize various kinds of nonequilibrium reservoirs. Active reservoirs refer to a special class of out-of-equilibrium reservoirs, which consist of a collection of self-propelled `active' agents. Examples of active agents range from microorganisms like bacteria, and macroscopic living entities like birds to artificially synthesized Janus particles and nanobots. The self-propelled nature of active particles leads to a range of intriguing features for systems coupled to active reservoirs. For example, single tracer particles suspended in an active bath shows modification of fluctuation-dissipation relation and equipartition theorem~\cite{seyforth2022nonequilibrium,wu,maes_activebath,ion_reservoir,khali2023does,Solon_2022,tracer_diff1,boriskovsky2024fluctuation,maggi,maggi_srep}, anomalous transport and non-Gaussian fluctuations~\cite{micromotor,kafri_greneck,transport_bacterial_turbulence,dhar2024active,Knezevic_2020,Bello2024,Put_2019} as well as force renormalizations~\cite{force_renormalization}.
Coupling with active reservoirs also leads to other interesting features including emergence of unusual thermodynamic properties~\cite{stirling_engine,D3SM01177A}, effective interaction among tracer particles~\cite{maggi_2,tracer_diff2}, sorting of polymer-like structures~\cite{muzzeddu2024migration} and capillary condensation~\cite{cappilary_condensation}.

A particularly important question is, how the nonequilibrium stationary state of an extended system is affected when driven by such active reservoirs. This question has recently been addressed in the context of energy transport through a harmonic chain, using a very simple model of an active reservoir\cite{activity_driven_chain,activity_stationary}. In these works, the effect of an active reservoir on a probe particle was modeled phenomenologically by introducing an `active' self-propulsion force, in addition to the usual dissipative and white-noise forces coming from an equilibrium thermal reservoir. 
This simple model showed several intriguing features of the NESS including negative differential conductivity and a non-trivial directional reversal of the active current. It was also shown that this current-carrying NESS cannot be generated in a system driven by thermal reservoirs with activity-dependent effective temperatures, despite having an activity-dependent local kinetic temperature in the bulk. 
In this model, the active noise kernel is taken to be independent of the dissipation coefficient, which is assumed to be a constant. Recent studies, however, have shown that this is not the case--- the noise and dissipation kernels arising from various microscopic models of active reservoirs, albeit violating FDR, are related to each other\cite{ion_reservoir,kafri_greneck,maes_activebath}. This raises an obvious and natural question--- how many of the unusual features exhibited by the minimal model of activity-driven NESS survive when one considers the effect of these non-trivial noise and dissipation kernels? A direct and systematic way to address this question is to consider an explicit microscopic model for the active reservoir,  the extended system, as well as the system-reservoir coupling. Perhaps the simplest model of an active reservoir is a one-dimensional chain of active particles, with nearest-neighbor interactions. While certain statistical properties of such active chains themselves have been studied recently \cite{paul2024dynamical,guptaactivechain,singh2021crossover,prakash2024tagged}, the role of such systems as active reservoirs have not been explored so far.

In this work, we propose a microscopic model of an active reservoir in the form of a one-dimensional chain of run-and-tumble particles (RTP)~\cite{rtp,rtp1} with nearest-neighbor interactions. In the absence of any interaction, a one-dimensional RTP shows a persistent motion with a dichotomous self-propulsion velocity; the activity of the particle is characterized by the persistence time.
We start with the simplest situation, where the active reservoir is a harmonic chain of such  RTPs, each of which has independent self-propulsion dynamics. The persistence time of all the reservoir particles is assumed to be the same, which characterizes the activity of the reservoir. The presence of activity results in a modified FDR, which we derive explicitly, by computing exactly the effective noise and dissipation kernels experienced by an inertial probe particle coupled linearly to one end of the reservoir chain.

We use these results to investigate the nonequilibrium stationary state and transport properties of an ordered harmonic chain, which is driven by two such active reservoirs at the ends with different activities $(\tau_1, \tau_N)$. We show that the presence of the activity drive introduces nontrivial spatial correlations in the system, unlike the thermally driven scenario, which we calculate exactly. In particular, in the stationary state, two characteristic length scales $\ell_{1,N} = \omega_c\tau_{1,N}$ emerge, where $\omega_c$ is the frequency of the harmonic chain, and velocities of the oscillators are correlated over a distance $\max(\ell_1,\ell_N)$.
We also compute the nonzero average energy current flowing through the harmonic chain due to the activity drive, which retains the negative differential conductivity (NDC) and current reversal as observed in Ref.~\cite{activity_driven_chain,activity_stationary}. Using numerical simulations, we show that our results remain qualitatively valid even when the reservoir particles have an anharmonic interaction. 

\section{Characterization of the active reservoir} \label{charectarization}

The behavior of a reservoir is usually characterized by its action on a probe particle coupled to it. Here we propose a simple model of an active reservoir as a one-dimensional ordered chain of active particles. In the absence of any interaction, the position $y(t)$ of a self-propelled active particle evolves via  an overdamped Langevin equation,
\bea
\nu \dot{y}=f(t),
\label{eq:leq1}
\eea
where $\nu$ is the friction coefficient and the stochastic force $f(t)$ models the self-propulsion. Different dynamics of $f(t)$ correspond to different active particle models \cite{AOUP,aoup1,rtp1,rtp,abp_fodor,abp}, the simplest one being the run-and-tumble particle (RTP), where $f(t)$ is a dichotomous noise that has a constant magnitude and changes sign intermittently. In general, the self-propulsion force is taken to be a stationary colored noise with zero mean, a characteristic time $\tau$, and an autocorrelation,
\bea
\la f(t)f(t') \ra =  h\left(t-t',\tau\right),\label{noise_correl_time}
\eea
where the functional form of $h(t)$ depends on the specific dynamics of $f(t)$. Note that, for any finite $\tau$, Eqs.~\eqref{eq:leq1} and \eqref{noise_correl_time} implies that the active particle dynamics automatically violates Fluctuation-Dissipation Theorem\cite{Kubo}.


\begin{figure}[ht]
\centering
\includegraphics[width=0.4\textwidth]{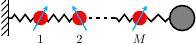}
\caption{Schematic representation of the active Rubin bath, consisting of $M$ overdamped active particles. The first particle is attached to a fixed wall while the last particle is coupled to a passive probe.}
\label{fig:single_particle}
\end{figure}
The active reservoir consists of $M$ such identical active particles with nearest-neighbor interaction mediated by a potential $V(z)$; see Fig.~\ref{fig:single_particle} for a schematic representation. We take a fixed boundary condition at one end---the left-most particle $l=1$ is attached to a fixed wall, while the other boundary particle $l=M$, is coupled to an inertial probe particle. The displacement $y_l$ of the $l$-th particle of the active reservoir from its equilibrium position evolves by,
\bea
\nu\dot{y}_l(t) &=& -\frac{\partial}{\partial y_l}[V(y_{l}-y_{l-1})+V(y_{l+1}-y_{l})]+{f_l}(t),~~ \forall\, l\in[1,M-1],\label{eom1}\\
\nu\dot{y}_M(t) &=& -\frac{\partial}{\partial y_M}[V(y_{M}-y_{M-1})+V(x_1-y_M)]+{f_M}(t),\label{eomM}
\eea
where $f_l(t)$ is the self-propulsion force on the $l$-th particle, which is assumed to be stationary with zero mean and autocorrelation 
\bea 
\la f_l(t) f_{l'}(t') \ra =\delta_{ll'}\,h(t-t',\tau). \label{eq:noise_llp}
\eea
Moreover, $x_1(t)$ denotes the displacement of the probe particle, which, in turn, evolves according to,
\bea
m \ddot{x}_1=-\frac{\partial}{\partial x_1}V(x_1-y_M). \label{eq:generic_prob}
\eea
For simplicity, we have taken the same interaction potential $V(z)$ between the right boundary particle and the probe. Note that, the fixed boundary condition for the first particle $l=1$ implies $y_0=0$.

The most direct way to characterize the behavior of a probe particle coupled to a reservoir is to write an effective equation of motion for it by integrating out the reservoir degrees of freedom. However, this is very hard for general reservoir models with arbitrary interaction and one needs to take recourse to approximate methods like infinite time-scale separation and perturbative techniques \cite{bhadra2016system,kruger2016modified}. A special case, where exact computations are possible, is when the couplings are linear in nature; examples include the Feynman-Vernon \cite{Feynman_Vernon}, and Rubin bath \cite{rubin_bath,quantum_brownian_rubin,das2012landauer} models to more recent models in the context of active particle dynamics \cite{ion_reservoir}. In this work, we adopt this approach and first consider a harmonic interaction potential $V(z)=\frac{\lambda}{2}z^2$. This leads to a set of linear equations of motion for the reservoir particles and the probe, 
\begin{align}
  \nu\dot{y}_l &= \begin{cases}
  \lambda({y}_{l+1}+{y}_{l-1}-2 {y}_l)+{f_l}(t),& \forall\, l\in[1,M-1],\\
    \lambda(x_1+{y}_{M-1}-2 {y}_M)+{f_M}(t), & \text{when }l=M,
   \end{cases}\label{eom_chain}\\
   \text{and}\quad
   m\ddot{x}_1&=\lambda({y}_{M}-x_1),\label{eom2}
\end{align}
with the boundary condition $y_0=0$. It should be noted that this model can be considered as an over-damped version of the Rubin bath with active noise. In the following, we characterize this \emph{active Rubin bath} by deriving the generalized Langevin Equation for the probe particle. 

To obtain an effective equation for $x_1(t)$, we need to solve Eq.~\eqref{eom_chain}, and express $y_M(t)$ in terms of $x_1(t)$. This can be done explicitly due to the linear nature of \eref{eom_chain} [see Appendix~\ref{detail_active_chain} for the details], which yields,
\begin{align}
 y_M(t) =\frac{ \lambda}{\nu} \int_{-\infty}^t ds \, x_1(s) \Lambda_{M  M}(t-s) + \frac{1}{\nu}\int_{-\infty}^t ds \sum_{k=1}^{M} \Lambda_{M k}(t-s)\, f_{k}(t),
 \label{bound_active}
\end{align}
where $\Lambda(z)$ is an $M \times M$ matrix with elements 
\bea
\Lambda_{j\ell}(z) = \frac 2{M+1} \sum_{k=1}^{M} \sin \frac{j k \pi}{M+1} \sin \frac{\ell k \pi}{M+1}e^{\mu_k z/\nu}~\text{with}~\mu_{k}=-4 \lambda \sin^2{\Big(\frac{k \pi}{2(M+1)}\Big)}.\label{eq:lambda_jl}
\eea 
Integrating the first term on the right-hand side of \eref{bound_active} by parts and substituting the resulting expression in Eq.~\eqref{eom2}, we get a generalized Langevin equation for the motion of the probe particle,
\begin{align}
m \ddot x_1(t) =  - \frac{\lambda \, x_1(t)}{(M+1)} - \frac{2\lambda}{M+1} \int_{-\infty}^t ds \, \dot x_1(s)  \sum_{k=1}^M \left(1+ \frac{\mu_k}{4 \lambda} \right) e^{\frac{\mu_k}{\nu}(t-s)} + \frac{\lambda}{\nu}\int_{-\infty}^t ds \sum_{k=1}^{M} \Lambda_{M k}(t-s)\,f_{k}(s). \label{eq:eff_eq_M}
\end{align}
It is useful to understand the physical significance of the various terms appearing in this effective equation. The first term on the right-hand side denotes the renormalized 
coupling constant of the probe with the reservoir. The second and third terms denote the dissipative and random forces experienced by the probe due to its coupling to the active reservoir.

We are particularly interested in the limit of large reservoir size $M$, where the effective coupling constant vanishes and we have  a simple form for the generalized Langevin equation,
\bea
m\ddot{x}_1=-\int_{-\infty}^{t} ds\,  \dot{x}_1(s)\gamma(t-s)+\Sigma(t),
\label{eq_eff1}
\eea
where the dissipation kernel 
\bea 
\gamma(t)  = \frac{2 \lambda}{M+1}\sum_{k=1}^M \cos^2{ \frac{k \pi}{2 (M+1)}}e^{\frac{\mu_k t}{\nu}},
\label{finiteM:gamma}
\eea 
and the effective noise 
\bea 
\Sigma(t) = \frac{2\lambda}{\nu(M+1)}\sum_{k=1}^{M} \sum_{j=1}^{M}(-1)^{j+1} \sin{\left(\frac{j \pi}{M+1}\right)}\sin{\left(\frac{kj \pi}{M+1}\right)}\int_{-\infty}^t ds\,e^{\frac{\mu_j}{\nu} (t-s)}\,f_{k}(s).
\label{finiteM:sigma}
\eea 

The behaviors of the dissipation kernel $\gamma(t)$ and the effective noise $\Sigma(t)$, in the thermodynamic limit, are discussed separately in the following. \\

\noindent{\bf Dissipation kernel:} In the thermodynamic limit $M \to \infty$, the summation over $k$ in the dissipation kernel \eref{finiteM:gamma} can be replaced by an integral over $u = \frac {k \pi}{2(M+1)}$, which leads to,
\bea 
\gamma (t) = \frac{4 \lambda}{\pi} \int_0^{\frac \pi 2} du \cos^2{u} \, \exp{\left[-\frac{4 \lambda t}{\nu}\sin^2{u} \right]}.
\eea 
This integral can be performed exactly, leading to a simple form for the dissipation kernel,
\bea
\gamma(t)=\lambda e^{-\frac{2 \lambda t}{\nu}}\Bigg[ I_0 \Bigg(\frac{2 \lambda t}{\nu}\Bigg)+I_1\Bigg(\frac{2 \lambda t}{\nu}\Bigg)\Bigg]\Theta(t),
\label{eq:diss:kernel}
\eea
where $\Theta(z)$ is the Heaviside-theta function and $I_n(z)$ denotes the $n$th order modified Bessel function of the first kind~\cite{DLMF}. 

Interestingly, for large $t\gg \nu/\lambda$, the dissipation kernel shows a power-law decay, $\gamma(t)\sim t^{-1/2}$. Such power-law decays are generic and have been observed in polymer chains and active baths\cite{power_law_kernel1,power_law_kernel2,power_law_kernel3,kafri_greneck}. Note that, in this system, the dissipation kernel \eref{eq:diss:kernel} depends only on the interaction potential and is independent of the self-propulsion force $f_l(t)$.

\begin{figure*}[t]
\centering
    \includegraphics[width=\textwidth]{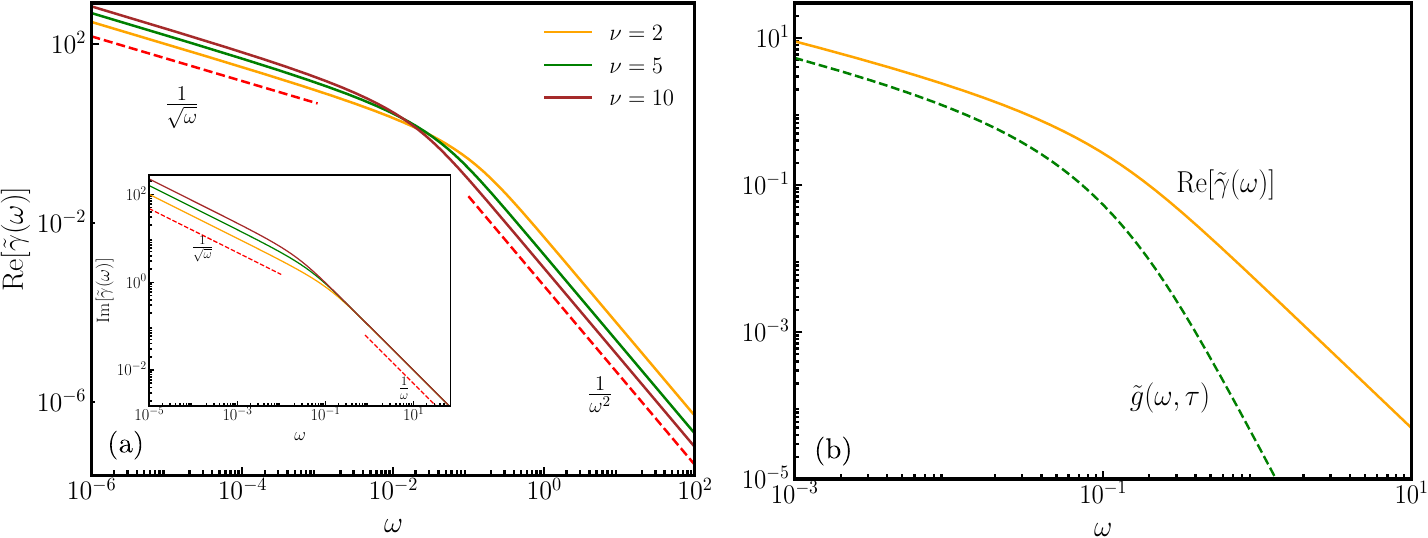} 

    \caption{Characterization of the active Rubin bath: (a) Plots of $\text{Re}[\tilde\gamma(\omega)]$ (main plot) and $\text{Im}[\tilde\gamma(\omega)]$ (inset) as functions of $\omega$ for $\lambda=0.1$, and different values of $\nu$ [see \eref{eq_memory_kernel}]. The red dashed lines indicate the asymptotic behaviors for small and large $\omega$. (b) Comparison of the reservoir spectra $\tilde g(\omega,\tau)$ and  $\text{Re}[\tilde{\gamma}(\omega)]$, illustrating the violation of FDR for active Rubin baths. Here we have used  $\lambda=0.1$, $\nu=2$ and $\tau=5$.}
\label{fig:analytical_memorykernel}
\end{figure*}

The spectral function of the reservoir, defined as the Fourier transform of the dissipation kernel $\tilde{\gamma}(\omega)=\int^{\infty}_0 dt \, e^{i \omega t}\gamma(t)$, plays an important role in determining the transport properties of a system driven by the reservoir. Clearly, for the real function $\gamma(t)$ given in Eq.~\eqref{eq:diss:kernel}, we must have $\mathrm{Re}[\tilde{\gamma}(-\omega)]=\mathrm{Re}[\tilde{\gamma}(\omega)]$ and $\mathrm{Im}[\tilde{\gamma}(-\omega)]=-\mathrm{Im}[\tilde{\gamma}(\omega)]$. Hence, it suffices to compute the spectrum for $\omega \ge 0$, which is given by,
\begin{align}
\tilde{\gamma}(\omega)=
-\frac{\nu}{2}\left[1+\sqrt{1+i \frac{4 \lambda}{\nu \omega}} \right]
={-\frac{\nu}{2}\Bigg[1-\sqrt{\sqrt{\frac{1}{4}+\frac{4 \lambda^2}{\nu^2 \omega^2}}+\frac{1}{2}}\Bigg]}
+i{ \frac{\nu}{2}\sqrt{\sqrt{\frac{1}{4}+\frac{4 \lambda^2}{\nu^2 \omega^2}}-\frac{1}{2}}}. \label{eq_memory_kernel}
\end{align}
For small $\omega$, both $\mathrm{Re}[{\tilde{\gamma}}]$ and $\mathrm{Im}[{\tilde{\gamma}}]$ decay as $\omega^{-1/2}$, consistent with the  large $t$ behaviour of $\gamma(t)$. On the other hand, for large $\omega$, $\mathrm{Re}[{\tilde{\gamma}}]$ and  $\mathrm{Im}[{\tilde{\gamma}}]$ decay as $\omega^{-2}$ and $\omega^{-1}$ respectively. Figure~\ref{fig:analytical_memorykernel}(a) illustrates these asymptotic behaviours of $\mathrm{Re}[{\tilde{\gamma}}]$ and $\mathrm{Im}[{\tilde{\gamma}}]$. \\

\noindent{\bf Noise autocorrelation:} We characterize the effective noise $\Sigma(t)$ defined in \eref{finiteM:sigma} by computing its mean and auto-correlation. Since the self-propulsion force is assumed to be a stationary process with a zero mean, we must have $\la\Sigma(t)\ra=0$. The autocorrelation of the effective noise $\Sigma(t)$ can be written using \eref{finiteM:sigma} and \eref{eq:noise_llp} as,
\begin{align}
\la \Sigma(t) \Sigma(t')\ra = \frac{4\lambda^2}{\nu^2(M+1)^2}\sum_{k=1}^{M} \sum_{j=1}^{M} \sum_{j'=1}^{M}(-1)^{j'+j} \sin{\left(\frac{j \pi}{M+1}\right)}\sin{\left(\frac{j' \pi}{M+1}\right)}\cr
\times \sin{\left(\frac{kj \pi}{M+1}\right)}\sin{\left(\frac{kj' \pi}{M+1}\right)}\int_{-\infty}^t ds\int_{-\infty}^{t'} ds'\,e^{\frac{\mu_j}{\nu} (t-s)}e^{\frac{\mu_{j'}}{\nu} (t'-s')}\,h(s-s',\tau),
\label{force_correl_1}
\end{align}
The sum over $k$ can be immediately performed using the identity $\sum_{k=1}^M \sin \frac{kj \pi}{(M+1)} \sin \frac {k j' \pi} {(M+1)} = \delta_{jj'} (M+1)/2$, which also allows us to perform the sum over $j'$. Finally, we arrive at, 
 \begin{align}
\la \Sigma(t) \Sigma(t')\ra = \frac{2\lambda^2}{\nu^2(M+1)} \sum_{j=1}^{M}  \sin^2{\left(\frac{j \pi}{M+1}\right)}
\int_{-\infty}^t ds\int_{-\infty}^{t'} ds'\,e^{\frac{\mu_j}{\nu} (t-s + t'-s')}\,h(s-s',\tau).
\end{align}
In the thermodynamic limit $M\to \infty$, the sum over $j$ can be replaced by an integral over $u = \frac {j \pi}{M+1}$, which leads to,
\begin{align}
\la \Sigma(t) \Sigma(t')\ra = \frac{2\lambda^2}{\nu^2}\int_0^{\pi} \frac{du}{\pi} \sin^2 u \, e^{\mu(u) (t+t')}  \int^{t}_{-\infty} ds \int^{t'}_{-\infty}  ds'e^{-\mu(u) (s+s')}h(s-s',\tau).\label{eq_left_effective_noise_correl}
\end{align}
where  
\bea
\mu(u)=- \frac{4 \lambda}{\nu} \sin^2{\frac u2}.
\eea
For our purpose, it is convenient to write down the noise autocorrelation in the frequency domain,
\bea
\la \tilde{\Sigma}(\omega) \tilde{\Sigma}(\omega')\ra=\int _{-\infty}^{\infty}dt \, e^{i \omega t}\int _{-\infty}^{\infty}   dt' \, e^{i \omega' t'}\la \Sigma(t) \Sigma(t')\ra.\label{noise_correl_freq}
\eea
Using \eref{eq_left_effective_noise_correl} in \eref{noise_correl_freq}, and performing the integrals over $t$ and $t'$, we get,
\bea
\la \tilde{\Sigma}(\omega) \tilde{\Sigma}(\omega')\ra 
=\frac{2\lambda^2 v_0^2}{\nu^2 \pi}\int_0^{\pi} du \sin^2 u 
 \int^{\infty}_{-\infty} ds \int^{\infty}_{-\infty}  \, ds'\frac{e^{i \omega (s-s')} e^{i (\omega+\omega') s'}}{(\mu(u)+i \omega)(\mu(u)+i \omega')} h(s-s',\tau).
\eea
The integrals over $s$ and $s'$ can also be performed exactly, leading to,
\bea
\la \tilde{\Sigma}(\omega) \tilde{\Sigma}(\omega')\ra= 2 \pi \delta(\omega+\omega')\tilde{g}(\omega),
\eea
with the effective noise spectrum of the active bath given by,
\bea
\tilde{g}(\omega)=\tilde{h}(\omega,\tau)\int^{\pi}_0 \frac{du}{2\pi} \frac{\sin^2 u}{ (1- \cos u)^2 + (\omega \nu / 2 \lambda)^2}.
\eea
Here $\tilde h(w,\tau)=\int_{-\infty}^{\infty}dt\,e^{i\omega t}h(t,\tau)$ denotes the spectrum of the active noise. Performing the integral over $u$, it turns out that,
\bea
\tilde g(\omega,\tau)  = \frac 1\nu \tilde h(\omega,\tau) \mathrm{Re}[\tilde{\gamma}(\omega)]\label{spectra_define},
\eea
where $\tilde{\gamma}(\omega)$ is given in \eref{eq_memory_kernel}. The above equation is one of the main results of this work and represents the modified FDR for the active Rubin bath. For an equilibrium bath at temperature $T$, consisting of passive oscillators, \eref{spectra_define} would reduce to the usual form of FDT $\tilde g(\omega,\tau)=T\, \mathrm{Re}[\tilde{\gamma}(\omega)]$. The temperature being replaced by a frequency-dependent function $\tilde h(\omega,\tau)$ indicates that, in general, such an active bath, with a given activity $\tau$, cannot be described by a unique effective temperature.

In what follows, we will mostly consider the case where the active oscillators follow a run-and-tumble dynamics, i.e., $f_l(t)$ is a dichotomous noise that alternates between $\pm v_0$ stochastically with a rate $(2\tau)^{-1}$. The corresponding the autocorrelation Eq.~\eqref{noise_correl_time} decays exponentially,
\bea
h(t,\tau)=v_0^2e^{-|t|/\tau}, \label{correl_rtp}
\eea
which in the frequency domain becomes a Lorentzian,
\bea
\tilde h(\omega,\tau)=\frac{2 v_0^2 \tau}{1+\omega^2 \tau^2}. \label{noise_rtp_freq}
\eea

\section{Harmonic chain driven by active Rubin baths}\label{model}
\begin{figure*}[th!]
\centering
\includegraphics[width=0.8\textwidth]{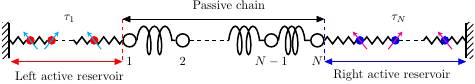}
\caption{Schematic representation of an ordered chain of $N$ harmonic oscillators driven by two active Rubin reservoirs of different activities $\tau_1$ and $\tau_N$.}
\label{fig:model}
\end{figure*}
In this section, we investigate the stationary state properties of a one-dimensional extended system, modeled by a harmonic chain, driven by two active Rubin baths defined in the previous section [see Fig.~\ref{fig:model}]. We consider a chain of $N$ oscillators, each of mass $m$, coupled with its nearest neighbors by a harmonic spring of stiffness $k$. The left and right boundary oscillators are coupled to two active reservoirs, each consisting of $M$ active particles, but with different activities $\tau_1$ and $\tau_N$, respectively. Let $\{ x_l; \,l \in [1, N] \},$ denote the displacement of the $l$-th oscillator from its equilibrium position. In the limit of thermodynamically large reservoirs $M \to \infty$, using the results of the previous section [see Eq.~\eqref{eq_eff1}], the equations of motion describing the time-evolution of $\{ x_l \}$ can be written as,
\bea
m\ddot{x}_1&=& k(x_2-x_1)-\int_{-\infty}^{t} ds\, \dot{x}_1(s)\, \gamma(t-s)+\Sigma_1(t),\cr
m\ddot{x}_l&=&k(x_{l-1}+x_{l+1}-2 x_l),~~~\forall\, l\in[2,N-1], \label{eom_passive}\\
m\ddot{x}_N&=&k(x_{N-1}- x_N)-\int_{-\infty}^{t} ds\, \dot{x}_N(s)\, \gamma(t-s)+\Sigma_N(t). \nonumber
\eea
Note that, for simplicity, we have assumed that the dissipation kernel $\gamma(t)$ is the same for the two reservoirs i.e., the friction coefficient $\nu$ and the coupling constant $\lambda$ are the same for both the reservoirs. However, the different activities of the reservoirs lead to different effective noises $\Sigma_1(t)$ and $\Sigma_N(t)$, which are independent of each other. This activity drive leads to a NESS of the harmonic chain which we characterize in the following.

We start by solving Eq.~\eqref{eom_passive}, which is most conveniently done by using a matrix notation, which recasts Eq.~\eqref{eom_passive} as,
\bea
\mathcal{M} \ddot{X}(t)=-\Phi X(t)-\int_{-\infty}^{t} ds \Gamma(t-s) \dot{X}(s)+\Xi(t), \label{eq_matrix_eom}
\eea
where $X(t)=(x_1(t),\cdots x_N(t))^T$, $\mathcal{M}$ is the mass matrix with elements $\mathcal{M}_{ij}=m \delta_{ij}$ and $\Phi$ is the tridiagonal coupling matrix with elements,
\bea
\Phi_{ij}=k(2\delta_{ij}-\delta_{i,j-1}-\delta_{i,j+1}-\delta_{1i}\delta_{1j}-\delta_{Ni}\delta_{Nj}). \label{phi}
\eea
The elements of the dissipation kernel matrix $\Gamma(t)$ and the noise vector $\Xi(t)$ are given by,
\bea
\Gamma(t)_{ij} = \gamma(t) (\delta_{i 1} \delta_{j 1}+ \delta_{i N}\delta_{j N}),\quad \text{and}\quad
\Xi_j(t) = \Sigma_1(t) \delta_{1j}+\Sigma_N(t) \delta_{Nj}.\label{kernel_matrix_noise_matrix}
\eea
Taking a Fourier transform, defined by $\tilde{X}(\omega)=\int_{-\infty}^{\infty} dt e^{i \omega t} X(t)$, of Eq.~\eqref{eq_matrix_eom}, we get an algebraic matrix equation in the frequency domain,
\bea
\tilde{X}(\omega)=G(\omega)\tilde{\Xi}(\omega), \label{fourier_sol}
\eea
where $\tilde{\Xi}(\omega)$ is the Fourier transform of $ {\Xi}(t)$. $G(\omega)$ is the Greens function matrix \cite{DharReview2008,Dhar2001,Transportbook} given by,
\bea 
G(\omega)=[-M\omega^2+\Phi-i\omega{\tilde{\Gamma}}]^{-1},\label{ap_tridiag}
\eea
where $G^{-1}(\omega)$ is clearly a tridiagonal matrix.

The displacement of the $l$-th oscillator, can be written from Eq.~\eqref{ap_tridiag} as, 
\begin{align}
x_l(t)=\int_{-\infty}^{\infty} \frac{d \omega}{2 \pi} e^{-i \omega t} [G_{l1}(\omega)\tilde{\Sigma}_1(\omega)+G_{lN}(\omega)\tilde{\Sigma}_N(\omega)],\label{position}
\end{align}
where the Fourier transforms of the effective noises $\tilde{\Sigma}_1(\omega)$ and $\tilde{\Sigma}_N(\omega)$ are given by [see Eq.~\eqref{spectra_define}],
 \bea
\la \tilde{\Sigma}_i(\omega) \tilde{\Sigma}_j(\omega')\ra = 2 \pi \delta_{ij} \delta(\omega+\omega')\, \tilde g(\omega,\tau_i),~~i,j=1,N.\label{correlation_freq}
\eea

Our goal is to characterize the NESS of the activity-driven harmonic chain by computing the stationary kinetic temperature profile $\la v_l^2 \ra$, two-time velocity autocorrelation of a single oscillator $\la v_l(0) v_l(t) \ra$, equal time velocity-velocity correlation $\la v_l v_{l'} \ra$ and the average current flowing through the system.
However, before going to the activity-driven case, we present a brief overview of the thermally driven scenario which will be useful to discern the effect of activity.

\subsection{Harmonic chain driven by thermal Rubin bath}\label{thermal_results}
The reservoir introduced in Sec.~\ref{charectarization} reduces to a thermal one at  temperature $T$ when the active noise $f_l(t)$ in Eq.~\eqref{eom_chain} is replaced by a white noise $\eta(t)$ with autocorrelation $\la\eta(t)\eta(t')\ra=2\nu T\delta(t-t')$. In this case, the effective noise spectrum and the dissipation kernel are related by the FDT,
\bea
\tilde g(\omega)  =  T\, \mathrm{Re}[\tilde{\gamma}(\omega)].\label{thermal_reservoir}
\eea
The nonequilibrium stationary state of a harmonic chain driven by two such thermal reservoirs has been studied extensively \cite{rllmothRieder,Dhar2001}. The resulting NESS is characterized by an energy current proportional to the temperature difference of the two reservoirs with temperature $T_1$ and $T_N$ respectively and a uniform kinetic temperature profile, given by the average temperature of the reservoirs,  $(T_1+T_N)/2$. In fact, the velocities of the bulk oscillators become uncorrelated in the thermodynamic limit i.e., 
\bea
\la v_l v_{l'} \ra= \frac{T_1+T_N}{ 2m} \delta_{l,l'}.\label{thermal_correl_spatial}
\eea 
Moreover, the two-time velocity correlation of a single oscillator in the bulk is given by,
\bea
\la v_l(t) v_l(0) \ra=\frac{T_1+T_N}{2m} J_0\left(2 \sqrt{\frac{k}{m}}\, t\right),\label{eq_thermal_two_p}
\eea
where $J_0(z)$ is the $0$-th order Bessel function of the first kind \cite{DLMF}.
Note that, although to the best of our knowledge, Eqs.~\eqref{thermal_correl_spatial} and \eqref{eq_thermal_two_p} have not been reported in this form, these come out as a result of a straightforward calculation, which is discussed later in Secs.~\ref{vel_vel_correl} and \ref{two_time_correl}. In the following, we investigate how the activity drive affects these observables.

\subsection{Stationary state correlations}
We start with the velocity correlation of the bulk oscillators. In general, the two-point velocity correlation of the $l$-th oscillator, 
 is given by using Eq.~\eqref{position} can be easily written as,
\begin{align}
\la v_l(t)v_{l'}(t') \ra =\sum_{i=1,N} \int_{-\infty}^{\infty} \frac{d\omega}{2 \pi} \omega^2 e^{-i \omega (t-t')}G_{li}(\omega)G_{l'i}^*(\omega)\tilde{g}(\omega,\tau_i)\label{eq_vel_vel_correl},
\end{align}
where we have used Eqs.~\eqref{position} and \eqref{correlation_freq}. 
To compute such correlations, we need the matrix elements $G_{li}(\omega)$, which can be computed explicitly owing to the tridiagonal structure of $G^{-1}(\omega)$ [see Appendix~\ref{stationary_correl_app}]. In particular,  the relevant elements for the calculation of the correlations are given by,
\begin{align}
G_{l1}(\omega) &= \frac{ \cos{(N-l)q}+\frac{c(\omega)}{2 k\sin{q}} \sin{(N-l)q} }
{c(\omega) \cos{(N-1)q}+d(\omega) \sin{(N-1)q} }, \text{and}~ G_{lN} (\omega)= G_{N-l+1,1}(\omega),
\end{align}
where $\omega$ and $q$ are related by 
\bea 
\omega=\omega_c \sin{\frac{q}{2}}, \quad \text{with} ~~ \omega_c = 2 \sqrt{\frac k m}.\label{omega_q_rel}
\eea 
Moreover, we have defined,
\begin{align}
c(\omega)=& 2\omega\, \mathrm{Im}[\tilde{\gamma}]  -m\omega^2 -2 i \omega \, \mathrm{Re}[\tilde{\gamma}] ,\\
d(\omega)=&\frac{\omega^2}{k \sin{q}}\left[\mathrm{Im}[\tilde{\gamma}]^2-\mathrm{Re}[\tilde{\gamma}]^2- m k \cos{q}-m\omega\,\mathrm{Im}[\tilde{\gamma}]  +i\mathrm{Re}[\tilde{\gamma}] \left(m \omega -2 \mathrm{Im}[\tilde{\gamma}]\right)\right],
\end{align}
for notational simplicity. We are particularly interested in the correlation among the bulk oscillators, i.e., $l,l' \ll N$, and in the thermodynamic limit $N\to \infty$, where the contribution to the integral 
\eref{eq:general_vel_correl} from frequency regime $|\omega|>\omega_c$ vanishes exponentially. Moreover, in this limit, one can integrate over the fast oscillations [see Appendix~\ref{stationary_correl_app} for details], which yields,
\begin{align}
     \la v_l(t)v_{l'}(t') \ra =\frac 1{\nu m}\sum_{i=1,N}\int_{0}^{\omega_c} \frac{d\omega}{2\pi}  e^{-i\omega(t-t')}\frac{\cos{[(l-l')q]}}{\sqrt{\omega_c^2- \omega^2}}\, \tilde{h}(\omega,\tau_i).\label{eq:general_vel_correl}
\end{align}
As expected, the spatio-temporal two-point correlation in the bulk is a function of the distance between the two oscillators $l-l'$, and time separation $t-t'$.
In the following, we separately discuss the equal-time spatial correlation $\la v_l v_{l'} \ra$, and the two-time correlation $\la v_l(0)v_l(t)\ra$ of a single oscillator in the bulk. 

\subsubsection{Velocity-velocity correlation}\label{vel_vel_correl}
\begin{figure*}[t]
\centering
    \includegraphics[width=0.5\textwidth]{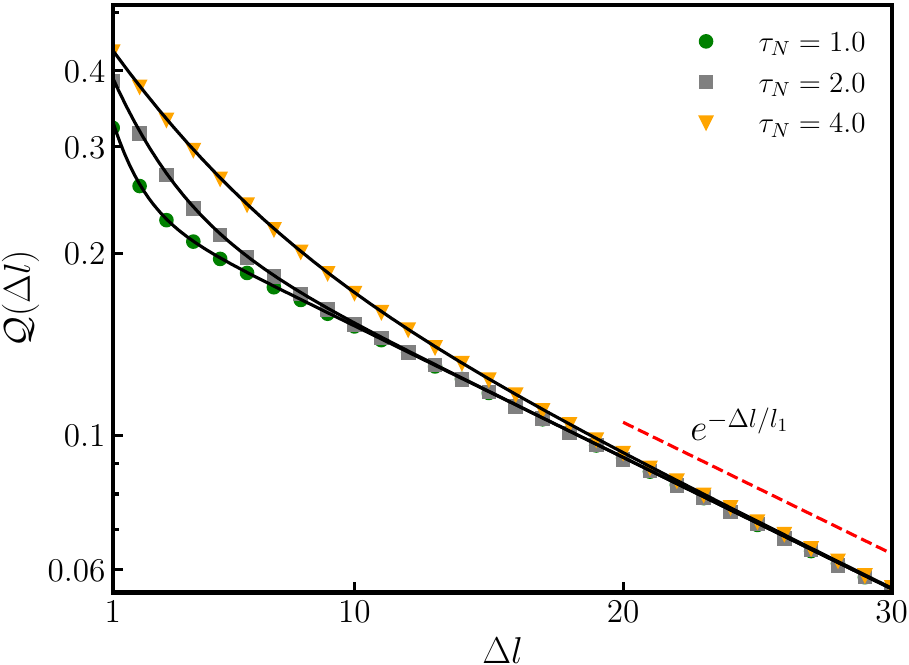} 
   \caption{Velocity-velocity correlation $\mathcal{Q}(\Delta l)$ {\it vs } $\Delta l = l'-l$ for a fixed $\tau_1=20$ and different values of $\tau_N$. The symbols denote the data obtained from numerical simulations with  $l=N/2$, bath size  $M=256$, and system size $N=256$.
   The black solid lines denote the analytical prediction Eq.~\eqref{eq_time_correl_rtp} and the red dashed line denotes the asymptotic exponential decay. Here we have taken $m=1=k=\nu=\lambda=v_0$.}
\label{fig:spatial_correl}
\end{figure*}
The velocity-velocity spatial correlation $\mathcal{Q}(l-l')\equiv\la v_l v_{l'} \ra$ can be obtained by putting $t=t'$ in \eref{eq:general_vel_correl},
\begin{align}
\mathcal{Q}(\Delta l) = \frac{v_0^2}{2 \pi \nu}\sum_{i=1,N}\int_{0}^{\pi} dq \frac{\tau_i \, \cos{(q\Delta l)}}{m+4 k \tau_i^2 \sin^2{\frac{q}{2}}},\label{spatial_correlmt}
\end{align}
where we have used \eref{omega_q_rel}. For $\Delta l\neq 0 $ the above integral is dominated by the contributions coming from the small $q$ regime and can be approximated as,
\bea
\mathcal{Q}(\Delta l) \approx \frac{v_0^2}{2 \pi \nu}\sum_{i=1,N}\int_{0}^{\infty} dq \frac{\tau_i \cos{(q\Delta l)}}{m+ k \tau_i^2 q^2}= \frac{v_0^2}{4\nu\sqrt{ k m}  }\sum_{i=1,N}\exp\left(-\frac{|\Delta l|}{\ell_i}\right),\label{eq_time_correl_rtp}
\eea
where $\ell_i =\tau_i \sqrt{k/m}$. 

Clearly, the active drive leads to the emergence of two characteristic length scales associated with the reservoirs, and velocities of the bulk oscillators are correlated over a separation $\max(\ell_1,\ell_N)$, determined by the reservoir with larger activity. The emergence of such a finite correlation is a direct consequence of the breaking of FDT and has been seen in the context of a boundary resetting-driven harmonic chain~\cite{resetting_chain} and is also expected to appear for simpler models of active reservoirs~\cite{activity_driven_chain,activity_stationary}. This is in sharp contrast to the thermally driven scenario, where the velocities of the bulk oscillators are uncorrelated [see  \eref{thermal_correl_spatial}]. The above prediction \eqref{eq_time_correl_rtp} is compared with the numerical simulations in Fig.~\ref{fig:spatial_correl} which shows an excellent agreement. 

\begin{figure*}[t]
\centering
    \includegraphics[width=\textwidth]{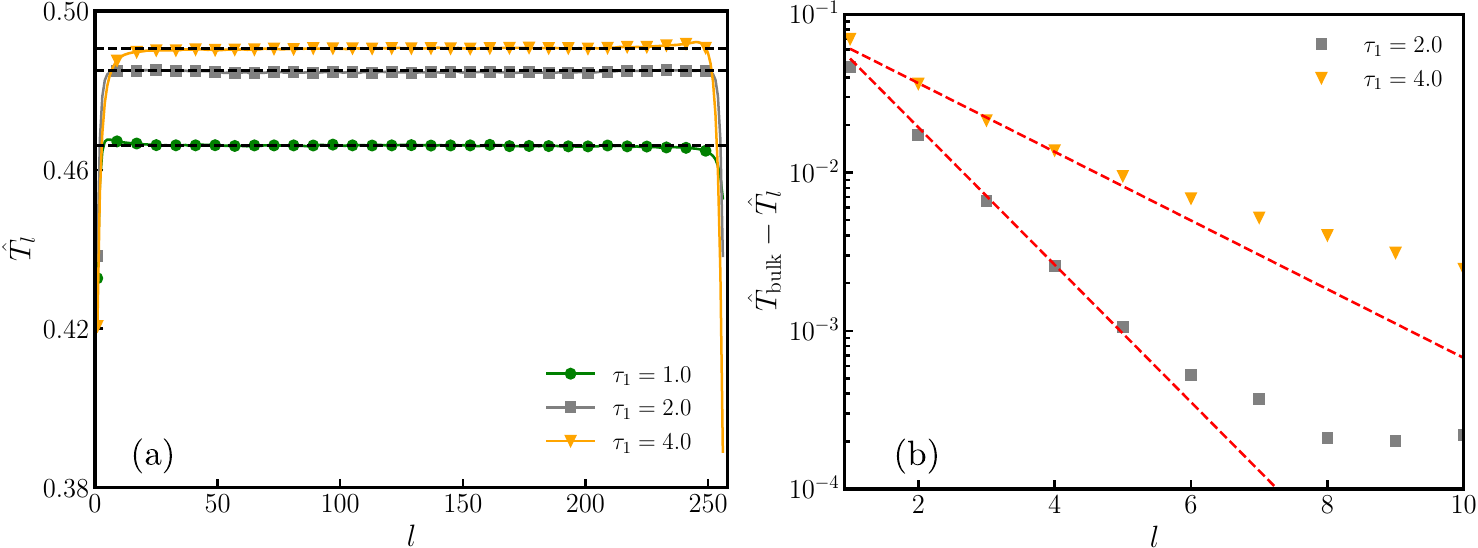} 

    \caption{(a) The kinetic temperature profile $\hat{T}_l$ for a fixed $\tau_N=2.0$ and different values of $\tau_1$.  The symbols indicate the data obtained from numerical simulations with $N=M=256$ and the dashed black lines indicate the predicted bulk temperature $\hat{T}_{\text{bulk}}$ [see \eref{eq:bulk_temp_rtp}]. (b) Plot of the deviation of the kinetic temperature profile from the bulk value near the left boundary, which exhibits an exponential decay, indicated by red dashed lines.
    The other parameters are $m=1=k=\nu=\lambda=v_0$.
    }
\label{fig:kin_temp_bulk}
\end{figure*}

\subsubsection{Kinetic temperature profile}

The kinetic temperature of the $l$-th oscillator $\hat{T}_l=m \la v_l^2(t) \ra$ is defined as the average kinetic energy in the steady state. From Eq.~\eqref{spatial_correlmt}, it is clear that, in the thermodynamic limit, the kinetic temperatures of the bulk oscillators attain a uniform value. This bulk kinetic temperature $\hat T_\mathrm{bulk}$, obtained by putting $\Delta l=0$ in \eref{spatial_correlmt}, is given by,  
\bea
\hat T_\text{bulk} = \frac{m v_0^2}{2 \pi \nu}\sum_{i=1,N}\int_{0}^{\pi} dq \frac{\tau_i }{m+ 4 k \tau_i^2 \sin^2{\frac{q}{2}}}=\frac{v_0^2}{2\nu } \sum_{i=1,N}\mathcal{T}(\tau_i), \quad \mathrm{where,}\quad \mathcal{T}(\tau)=\frac{ \tau}{\sqrt{1+\frac{4 k}{m}\tau^2}}. ~~~\label{eq:bulk_temp_rtp}
\eea
It is noteworthy that the bulk kinetic temperature does not depend on the dissipation kernel, and is determined only by the activity of the reservoirs. In fact, the form of $\hat T_\text{bulk}$ is the same obtained in Ref.~\cite{activity_driven_chain}, where the active reservoir was modeled by a single correlated force. In fact, it has also been shown that, although the form of \eref{eq:bulk_temp_rtp} is tempting [see Sec.~\ref{thermal_results}] to associate an effective temperature $v_0^2 \mathcal{T}(\tau_i)/\nu$ to the $i$-th active reservoir, such a picture does not capture the effect of activity, except in the passive limit $\tau \to 0$. In this limit, $\mathcal{T}(\tau)\simeq \tau$, and the bulk temperature can be expressed as,
\bea
\hat{T}_\text{bulk}=\frac{T_1^\mathrm{eff}+T_N^\mathrm{eff}}{2}\quad \mathrm{with}\quad T_i^\mathrm{eff}= \frac{v_0^2 \tau_i}{\nu}. \label{effective_temp}
\eea

Figure~\ref{fig:kin_temp_bulk}(a) shows the kinetic temperature profile for different values of $(\tau_1, \tau_N)$ along with the analytic prediction  \eref{eq:bulk_temp_rtp}. As expected, for any finite chain, $\hat{T}_l$ deviates from the $\hat{T}_\text{bulk}$ near the two boundaries, i.e., for $l\ll N$ and $l\sim N$.
Figure~\ref{fig:kin_temp_bulk}(b) illustrates that these boundary layers decay exponentially.

\subsubsection{Two-time velocity correlation}\label{two_time_correl}

\begin{figure*}[t]
\centering
    \includegraphics[width=0.5\textwidth]{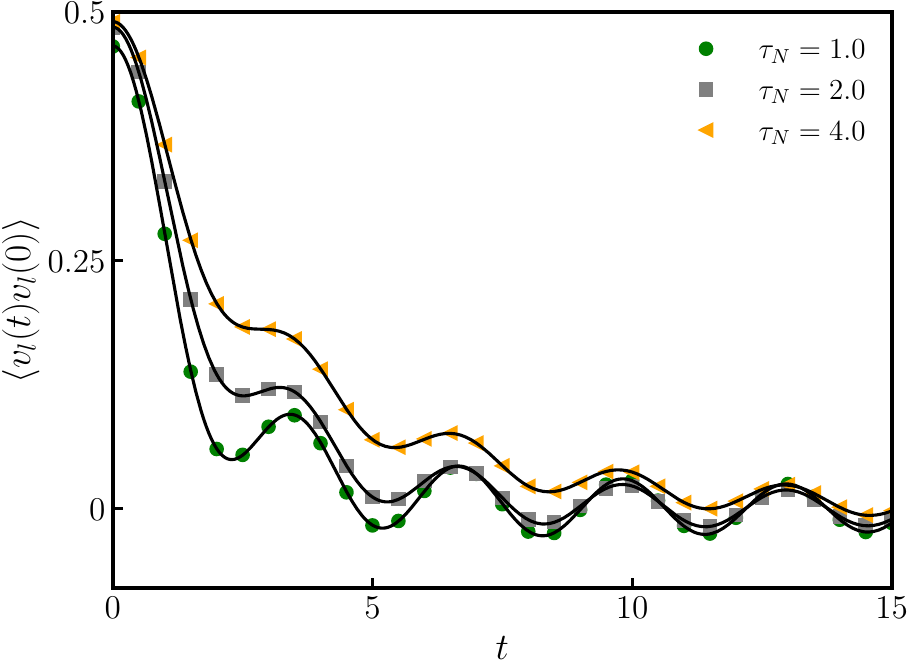} 

    \caption{Temporal decay of the velocity autocorrelation $\la v_l(t)v_{l}(0) \ra$ for a bulk oscillator with a fixed $\tau_1=2.0$ and different values of $\tau_N$. The symbols indicate the data obtained from numerical simulations with $l=N/2$  and $N=M=256$, and the solid lines indicate the analytical prediction \eref{two_time_general_driving_time_correl}. The other parameters are given by $m=1=k=\nu=\lambda=v_0$.
    }
\label{fig:temporal_correl_integral}
\end{figure*}

Next, we focus on the stationary two-time velocity autocorrelation of a single oscillator, which can be obtained by putting $l=l'$ in Eq.~\eqref{eq:general_vel_correl}. Using Eqs.~\eqref{omega_q_rel} and \eqref{noise_rtp_freq}, it is most conveniently expressed as, 
\bea
\la v_l(t)v_{l}(0) \ra=\frac{v_0^2}{2 \pi \nu}\sum_{i=1,N}\int_0^{\pi} dq \frac{\tau_i\,\cos{\left(\omega_c t \sin{\frac{q}{2}} \right)}}{m+4 k \tau_i^2 \sin^2{\frac{q}{2}}}.\label{two_time_general_driving_time_correl}
\eea
To evaluate the $q$-integral, we use the variable transformation $z=\omega_c \sin (q/2)$, which recasts \eref{two_time_general_driving_time_correl} as,
\bea
 \la v_l(t)v_{l}(0) \ra=\frac{v_0^2}{ \nu m}\sum_{i=1,N}\int_{0}^{\omega_c}\frac{d z}{\pi} \frac{\tau_i \cos z t}{\sqrt{\omega_c^2-z^2}(1+\tau_i^2 z^2)}.\label{large_time_rtp}
\eea 
The above integral can be numerically evaluated to obtain the two-time velocity correlation at all times. Figure~\ref{fig:temporal_correl_integral} shows the temporal decay of $\la v_l(t) v_l(0) \ra$ for different values of activity drive. The oscillatory nature of the two-time correlation is qualitatively similar to the thermally driven case [see \eref{eq_thermal_two_p}] and it is useful to investigate the effect of activity quantitatively.
To this end, we evaluate the integral in \eref{large_time_rtp} term by term by expanding $(1+\tau_i^2 z^2)^{-1}$ in a power series of $\tau_i$, which leads to, 
\begin{align}
\la v_l(t)v_{l}(0) \ra&=\frac{v_0^2}{ \nu m \pi}\sum_{i=1,N}\tau_i\sum_{n=0}^{\infty}(-\tau_i^2)^{n} \int_{0}^{\omega_c}d z \frac{ z^{2 n} \cos z t}{\sqrt{\omega_c^2-z^2}} \cr
&=\frac{v_0^2}{2 \nu m}\sum_{i=1,N}\tau_i\left[ \sum_{n=0}^{\infty} \left(-\frac{4k\tau_i^2}{m}\right)^n \Gamma \left(n+\frac{1}{2}\right){}_1 \tilde F_2\left[n+\frac 12; \frac 12,n+1;-\frac{k t^2}{m} \right]\right].~~~~~ \label{expand_vel_corr}
\end{align} 
Here ${}_1 \tilde F_2(a_1; b_1, b_2;z)$ denotes the regularized generalized Hypergeomatric function \cite{DLMF}. To understand the effect of activity on two-time velocity correlation, it is useful to analyze the asymptotic behavior of $\la v_l(t) v_l(0) \ra$ in the short-time ($t \ll \omega_c^{-1}$) and long-time ($t \gg \omega_c^{-1}$) regimes.

\begin{figure*}[t]
\centering
    \includegraphics[width=\textwidth]{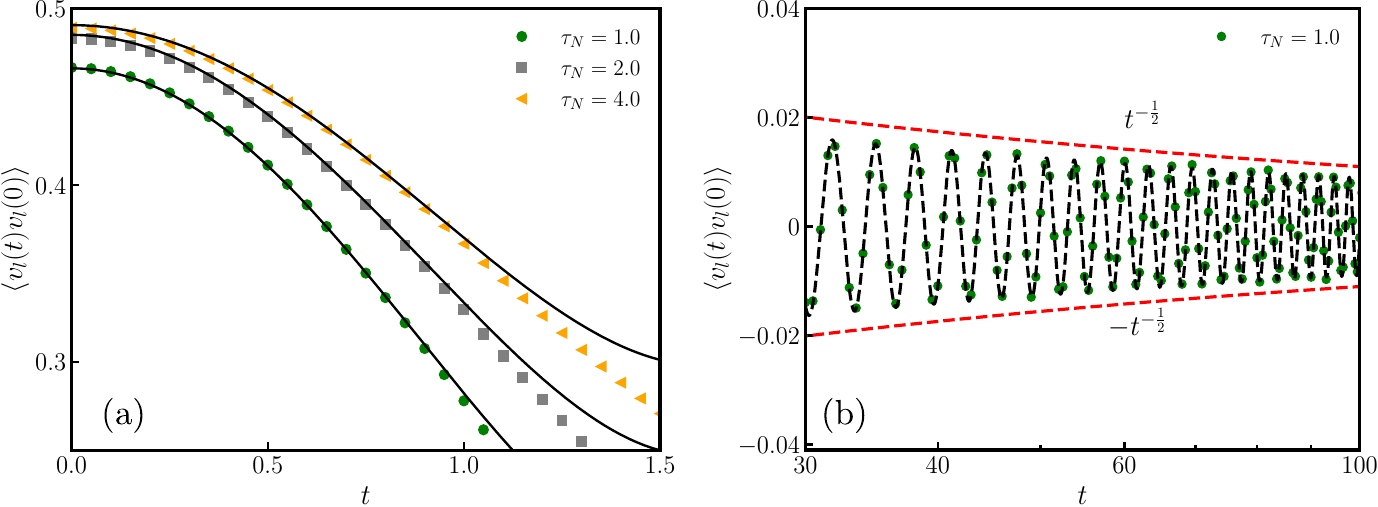} 

    \caption{Asymptotic behaviour of $\la v_l(t)v_{l}(0) \ra$ in (a) the short-time $(t \ll 1/\omega_c)$ and (b) the long-time $(t \gg 1/\omega_c)$ regimes. In both cases, we have taken $\tau_1=2$. The solid lines in (a) indicate the analytical prediction \eref{vel_correl_rtp_shorttime} while the dashed line in (b) corresponds to the analytical prediction \eref{eq_time_part_large}. The red dashed line 
    marks the $1/\sqrt{t}$ envelop of the Bessel function.
    The symbols show the same data used in Fig~\ref{fig:temporal_correl_integral}. }
\label{fig:temporal_correl_shorttime}
\end{figure*}

To extract the short-time behavior of $\la v_l(t) v_l(0) \ra$ we first expand ${}_1\tilde F_2(a_1;b_1,b_2,-z)$ for small values of $z$ \cite{DLMF},
\begin{align}
{}_1\tilde F_2\left[n+\frac 12; \frac 12,n+1;-z \right]=\frac{1}{\sqrt{\pi } n!}-\frac{ z (2n+1) }{ \sqrt{\pi } (n+1)!}+\frac{ z^2\left(2n+1\right) \left(2n+3\right) }{6 \sqrt{\pi } (n+2)!}+O(t^{6}).
\end{align}
Substituting the above equation in \eref{expand_vel_corr} and performing the sum over $n$, we get, 
\begin{align}
\la v_l(t)v_{l}(0) \ra=
\frac{v_0^2}{2 \nu m}\sum_{i=1,N}\Biggl[\mathcal{T}(\tau_i)-\frac{t^2}{2 \tau_i^2}\big(\tau_i-\mathcal{T}(\tau_i)\big)-\frac{t^4}{24 \tau_i^4}\left(\tau_i-\mathcal{T}(\tau_i)-\frac{2 k \tau_i^3}{m} \right)\Biggr]+O(t^{6}),\label{vel_correl_rtp_shorttime}
\end{align} 
where $\mathcal{T}(\tau)$ is defined in \eref{eq:bulk_temp_rtp}. Note that, as expected, in the $t\to 0$ limit, $\la v_l(t) v_l(0) \ra$ converges to the bulk kinetic temperature [see \eref{eq:bulk_temp_rtp}]. 
The short-time behavior of the two-time correlation is illustrated in Fig.~\ref{fig:temporal_correl_shorttime}(a). It is noteworthy that the anomalous short-time behavior \eref{vel_correl_rtp_shorttime}, which is qualitatively different than the same in the thermally driven scenario [see \eref{eq_thermal_two_p}], shows strong signatures of activity.

To obtain the large-time behavior of $\la v_l(t) v_l(0)\ra$, we note that for large $z\gg 1$,
\begin{align}
\Gamma \left(n+\frac{1}{2}\right){}_1\tilde F_2\left[n+\frac 12; \frac 12,n+1;-z \right]\approx
J_0\left(2 \sqrt{z}\right)
=\frac{  \sin \left(2\sqrt{z} \right)+ \cos \left(2\sqrt{z}\right)}{ \sqrt{2\pi } z^{1/4} }+O\left(\frac{1}{z^{3/4}}\right).\label{hyper_expansion}
\end{align}
Using the above equation in \eref{expand_vel_corr} we get the large-time behavior $t \gg \omega_c$ of the two-time velocity correlation $\la v_l(t) v_l(0)\ra$,
\begin{align}
  \la v_l(t)v_{l}(0) \ra \simeq \frac{v_0^2}{2 \nu m}J_0\left(2 t\sqrt{\frac{k}{m}}\right)\sum_{i=1,N} \sum_{n=0}^{\infty}\tau_i \left(-\frac{4k\tau_i^2}{m}\right)^n
= \frac{v_0^2}{2 \nu m}J_0\left(2 t\sqrt{\frac{k}{m}}\right)\sum_{i=1,N} \frac{ \mathcal{T}^2(\tau_i)}{ \tau_i} ,\label{eq_time_part_large}
\end{align}
where $\mathcal{T}(\tau)$ is defined in \eref{eq:bulk_temp_rtp}. The large time behavior of $\la v_l(t)v_{l}(0) \ra$ is shown in  Fig.~\ref{fig:temporal_correl_shorttime}(b), which illustrates its oscillatory decay with a $1/\sqrt{t}$ envelop.

It is noteworthy that \eref{eq_time_part_large} is similar to \eref{eq_thermal_two_p}, i.e., the velocity two-time correlation in the thermally driven scenario, but with a prefactor different from the bulk kinetic temperature $\hat T_\mathrm{bulk}$. This provides additional evidence that $\mathcal{T}(\tau)$  can not be thought of as an effective temperature for the active reservoirs, in general.  However, as mentioned before, a consistent effective temperature picture arises in the passive limit $(\tau_1, \tau_N) \ll \sqrt{k/m}$, where \eref{eq_time_part_large} resembles \eref{effective_temp} with $T_i^\mathrm{eff}$ playing the role of the effective temperature of the $i$-th bath.

\subsection{Stationary state current}\label{stat_current_harmonic}

The active reservoirs coupled to the boundary oscillators are expected to drive an energy current through the system when $\tau_1 \ne \tau_N$. To compute this current, it suffices to consider the instantaneous work done by one of the reservoirs (say, the left one) on the corresponding boundary oscillator. Thus, in the stationary state, the average energy current is given by~\cite{Transportbook,DharReview2008,Dhar2001},
\bea
J_\text{act}=\Big\la \Big(-\int_{-\infty}^{t} ds \, \dot{x}_1(s)\gamma(t-s)+\Sigma_1(t)\Big) \dot{x}_1(t)\Big\ra.
\label{eq_current_part}
\eea

\begin{figure}[t]
\centering
    \includegraphics[width=\textwidth]{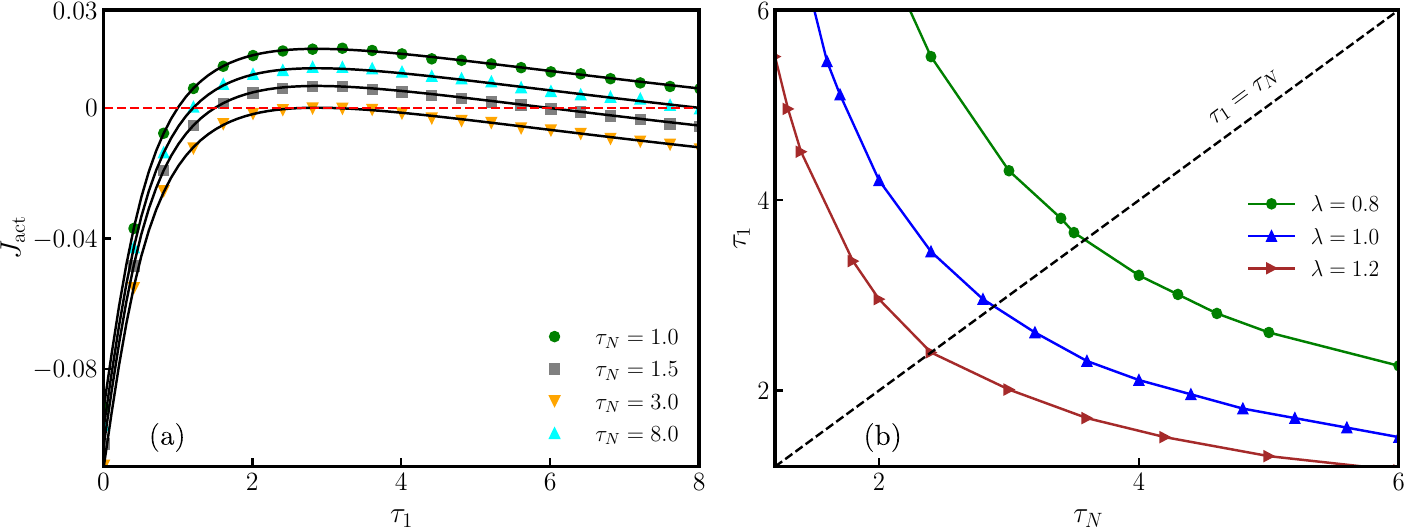} 
    \caption{(a) Plot of average energy current $J_\text{act}$ as a function of $\tau_1$ for different $\tau_N$. The black solid lines are obtained by performing the integral in \eref{average_energy_current} numerically. The symbols indicate the data obtained from numerical simulations with a system size $N=256$ and bath size $M=256$ and $m=1=k=\nu=\lambda=v_0$. Each curve crosses zero twice--- at $\tau_1=\tau_N$, and a non-trivial point $\tau_1^*(\tau_N)$. (b) Current reversal in the $(\tau_1,\tau_N)$ plane: the dashed black line indicates the trivial reversal line $\tau_1=\tau_N$ and the symbols indicate the nontrivial current reversal curves $\tau_1^*(\tau_N)$ for different values of $\lambda$.
    }
\label{fig:average_current}
\end{figure}

This average active current $J_\mathrm{act}$ can be computed using the Green's function formalism introduced in Ref.~\cite{Dhar2001} and adapted for nonequilibrium baths in Ref.~\cite{activity_driven_chain,activity_stationary}. The details of the computation are provided in Appendix~\ref{appendix_formalism}; here we quote the main result. The active current is given by a `Landauer-like' formula,
\begin{align}
J_\text{act}=\int_{-\infty}^{\infty} \frac{d \omega}{2 \pi} \omega^2 \left|G_{1N}\right|^2 \mathrm{Re}[\tilde{\gamma}(\omega)]^2\left[\tilde{h}(\omega,\tau_1)-\tilde{h}(\omega,\tau_N)\right],\label{eq_average_cur}
\end{align}
where the matrix $G(\omega)$, dissipation kernel $\tilde\gamma(\omega)$ and the autocorrelation of the active noise $\tilde h(\omega,\tau)$ are defined in Eqs.~\eqref{ap_tridiag}, \eqref{eq_memory_kernel} and \eqref{noise_rtp_freq}, respectively. Note that, the presence of frequency-dependent function $\tilde h(\omega, \tau)$ in \eref{eq_average_cur}, which is indicative of the violation of FDR of the active reservoir, distinguishes the above expression from the case of the equilibrium bath [see Eq.~(67) of Ref.~\cite{DharReview2008}].

We are particularly interested in the thermodynamic limit $N\to \infty$, where \eref{eq_average_cur} reduces to [see Appendix~\ref{appendix_formalism}],
\bea
J_\text{act}=\int_0^{\omega_c}\frac{d \omega}{4\pi} \frac{\mathrm{Re}[\tilde{\gamma}]\sqrt{m(4 k -m \omega^2)}}{(m k +|\tilde{\gamma}|^2- \mathrm{Im}[\tilde{\gamma}] m \omega)}\left[\tilde{h}(\omega,\tau_1)-\tilde{h}(\omega,\tau_N)\right]\label{average_energy_current}.
\eea 
Although the $\omega$ integral in the above equation can not be performed exactly to obtain a closed form for $J_\text{act}$, it can be evaluated numerically to arbitrary accuracy for any values of $(\tau_1, \tau_N)$. This is illustrated in Fig.~\ref{fig:average_current}(a) where we have plotted the analytical prediction \eref{average_energy_current} with $J_\text{act}$ measured from numerical simulations which shows an excellent agreement.

From Fig.~\ref{fig:average_current}(a) it is apparent that the active current shows a non-monotonic behavior as a function of $\tau_1$ as well as a non-trivial direction reversal at $\tau_1=\tau_1^*(\tau_N)$. This reversal point $\tau_1^*$ depends on reservoir coupling strength $\lambda$ which is illustrated in Fig.~\ref{fig:average_current}(b) where $\tau_1^*$ is plotted as a function of $\tau_N$ for three different values of $\lambda$. As the average current $J_\text{act}$ is a nonmonotonic function of $\tau_1$, the differential conductivity $\frac{d J_\text{act}}{d \tau_1}<0$ for a range of $\tau_1$. The negative differential conductivity and nontrivial direction reversal are also reported in Ref.~\cite{activity_driven_chain} using a much more simplified version of the active reservoir. The emergence of these features, even for the microscopic model of the active reservoir, indicates that these behaviors are rather robust which we illustrate in the following section using a more generalized model of the active reservoir.

\section{Generalizations to non-Markovian and non-linear reservoirs}\label{robuset_result}

The linear nature of the active chain and the Poissonian tumbling protocol makes the active Rubin model analytically treatable. An obvious question is whether the results obtained so far are special due to the simplicity of this model. In this section, we investigate this question by generalizing the model of the active reservoir by introducing a Non-markovian tumbling protocol and non-linear interactions among the reservoir particles. It turns out that the qualitative behavior of the NESS remains the same in both of these cases, which illustrates the robustness of our results.

\subsection{Non-Markovian tumbling protocol}

In general, the waiting time between two consecutive tumblings of the RTPs can be drawn from a distribution $\mathcal{P}(t,\tau)$. The constant rate Markovian protocol considered so far corresponds to the case where the waiting-time distribution is exponential i.e., $\mathcal{P}(t,\tau)=\exp(-t/\tau)/\tau$. One of the simplest ways to introduce a non-Markovian flipping protocol is to consider a Gamma-distribution $\mathcal{P}(t,\tau)=(t/\tau^2)\exp(-t/\tau)$ for the waiting time, where $\la t \ra=\tau$ still characterizes the activity. The corresponding autocorrelation of the active force $f_l(t)$ in the time as well as in the frequency domain are given by~\cite{clock_model1,psf_twostate},
\bea
h(t,\tau)= v_0^2 \exp{(-|t|/\tau)} \cos ( t/\tau),\quad \mathrm{and}\quad \tilde h(\omega,\tau)=\frac{2v_0^2 \tau (2+\omega^2 \tau^2)}{4+\tau^4\omega^4}, \label{noise_clock_freq}
\eea 
respectively. In this section, we discuss how the NESS of the activity-driven harmonic chain changes when the reservoir particles follow this particular flipping protocol.

\begin{figure*}[t]
\centering
    \includegraphics[width=\textwidth]{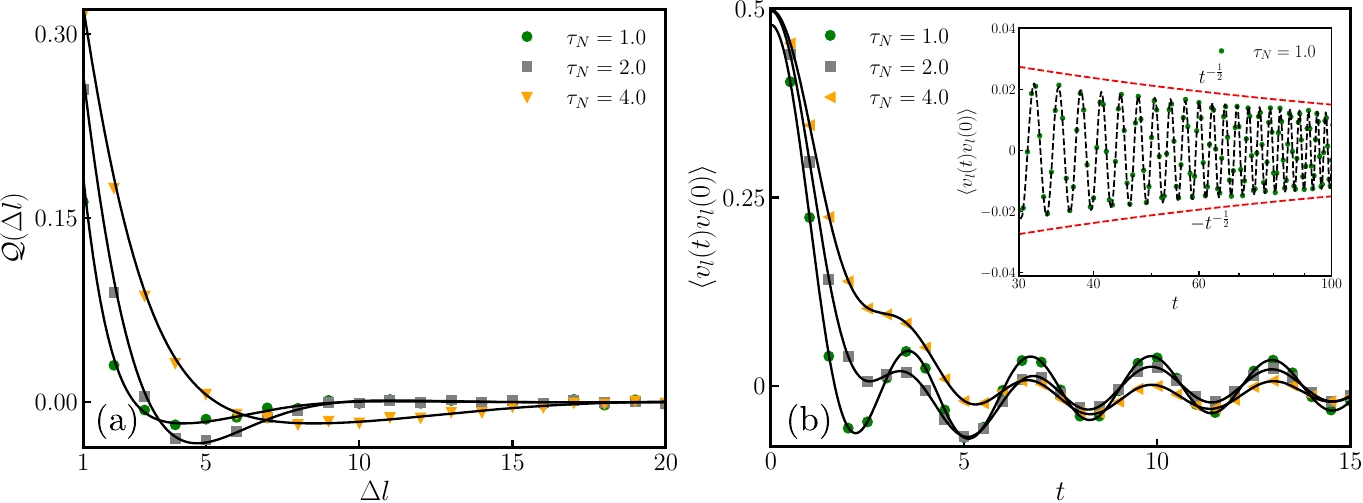} 

    \caption{Spatial and temporal velocity correlations for non-Markovian tumbling protocol in the reservoir: (a) Plot of $\mathcal{Q}(\Delta l)$ {\it vs} $\Delta l=l'-l$ for $l=N/2$, $\tau_1=2.0$ and different values of $\tau_N$. The black solid lines denote the analytical prediction Eq.~\eqref{eq_time_correl_clock}, while the symbols correspond to the numerical simulations. (b) Plot of the temporal velocity correlation of the middle oscillator $l=N/2$, for $\tau_1=2.0$ and different values of $\tau_N$. The analytical predictions are indicated by solid black lines [see \eref{two_time_clock_driving_time_correl}] while the symbols denote the numerical simulations. The inset illustrates the long-time asymptotic behavior predicted in \eref{eq:longtime_nopoissonian} for the curve corresponding to $\tau_N=1$ in the main plot. The parameters used in the numerical computation are as mentioned in fig.~\ref{fig:temporal_correl_integral}.
   }
\label{fig:temporal_spatial_nonmarkov}
\end{figure*}

The two-point velocity correlation of the bulk oscillator $\mathcal{Q}(\Delta l)$, in this case, can be obtained by substituting Eq.~\eqref{noise_clock_freq} and $t=t'$ in Eq.~\eqref{eq:general_vel_correl}. The integration can be performed exactly and yields,
\bea
\mathcal{Q}(\Delta l) =\frac{v_0^2}{\nu\sqrt{8 k m}  }\sum_{i=1,N}\exp\left(-\frac{|\Delta l|}{\ell_i}\right)\cos\left(\frac{\Delta l}{\ell_i}\right)\quad \mathrm{where}\quad\ell_i =\tau_i \sqrt{k/m}.\label{eq_time_correl_clock}
\eea
Clearly, in this case, too, we have the emergence of the two active length scales $(\ell_1,\ell_N)$ which remain the same as in the constant rate flipping case. However, the exponential decays are modulated by an oscillator function which makes it allows negative values for the correlation $\mathcal{Q}(\Delta l)$ for some values of $\Delta l$. Fig.~\ref{fig:temporal_spatial_nonmarkov}(a) illustrates the oscillator behavior of $\mathcal{Q}(\Delta l)$ for different values of the activity drive.

The two-time velocity correlation of a single oscillator $ \la v_l(t)v_{l}(0) \ra$ for the non-Markovian flipping protocol can also be obtained by using Eq.~\eqref{noise_clock_freq} in Eq.~\eqref{eq:general_vel_correl} and substituting $l=l'$ and $t'=0$. This leads to,
\bea
 \la v_l(t)v_{l}(0) \ra=\frac{v_0^2}{ \nu m}\sum_{i=1,N}\int_{0}^{\omega_c}\frac{d z}{\pi} \frac{\tau_i \cos z t}{\sqrt{\omega_c^2-z^2}}\frac{\left(2 +\tau_i^2 z^2\right)}{\left(4+\tau_i^4 z^4\right)}\quad \mathrm{with}\quad \omega_c=2 \sqrt{\frac{k}{m}},\label{two_time_clock_driving_time_correl}
\eea 
which can be evaluated numerically for all time. In the long-time limit $t\gg 1/\omega_c$, the dominant contribution to the integral in \eref{two_time_clock_driving_time_correl} comes from the region $z\simeq \omega_c$ and one can get a closed form expression, 
\bea
\la v_l(t)v_{l}(0) \ra \simeq J_0\left(2 t\sqrt{\frac{k}{m}}\right) \sum_{i=1,N} \frac{v_0^2 \tau_i  }{ 2 \nu m }\frac{\left( 2+ \omega_c^2 \tau_i^2\right)}{\left(4+\omega_c^4 \tau_i^{4}\right)}.\label{eq:longtime_nopoissonian}
\eea
The temporal decay of $\la v_l(t)v_{l}(0) \ra$ for the non-Markovian tumbling protocol is shown in fig.~\ref{fig:temporal_spatial_nonmarkov}(b). 

\begin{figure*}[ht]
\centering
    \includegraphics[width=\textwidth]{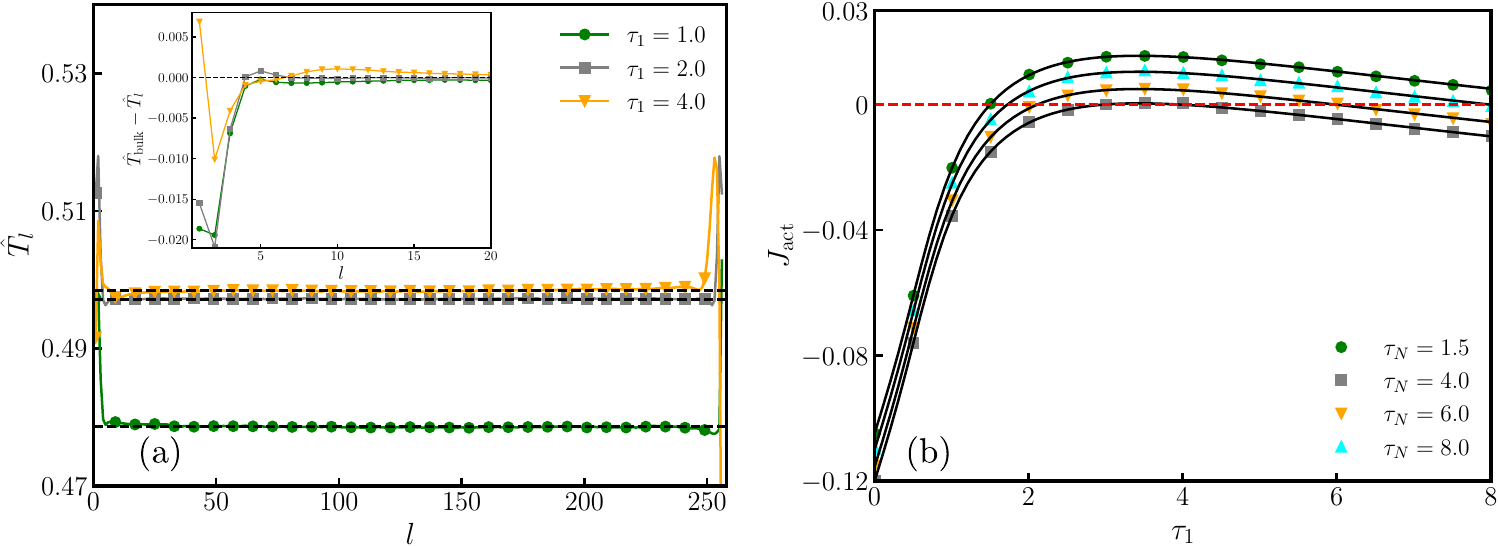} 

    \caption{ Transport properties for non-Markovian tumbling protocol in the reservoirs: (a) Kinetic temperature profile $\hat{T}_l$ for $\tau_N=2.0$ and different values of $\tau_1$. The predicted value of $\hat{T}_\text{bulk}$ in \eref{kin_temp_clock}, denoted by black dashed lines, is shown along with the data obtained from numerical simulations, indicated by symbols. The inset illustrates the oscillatory behavior of the boundary layer near $l=1$.
    (b) Plot of $J_\text{act}$ {\it vs} $\tau_1$ for different values of $\tau_N$. The symbols and the black lines indicate the analytical prediction \eref{average_energy_current} and the data obtained from numerical simulations, respectively.  The simulations are performed on a system of size $N=256$ with bath size $M=256$ and  $m=1=k=\nu=\lambda=v_0$.
    }
\label{fig_clock_current}
\end{figure*}

It is also straightforward to calculate the bulk kinetic temperature, which, in this case, turns out to be,
\bea
\hat{T}_\mathrm{bulk} =\frac{v_0^2}{4 \nu}\sum_{i=1,N}\frac{  \tau_i \left(m+2 k \tau_i^2+\sqrt{m^2+4 k^2 \tau_i^4}\right)}{  \sqrt{\left(2m^2 +8 k^2 \tau_i^4 \right)\left(1+\sqrt{1+\frac{4 k^2}{m^2} \tau_i^4}\right)}}.\label{kin_temp_clock}
\eea 
Figure~\ref{fig_clock_current}(a) shows the kinetic temperature profile $\hat T_l = m \la v_l^2 \ra $ for different values of $\tau_1$. Interestingly, the non-Markovian flipping protocol gives rise to an oscillatory behavior in the boundary layer, which is illustrated in the inset of Figure~\ref{fig_clock_current}(a). In the passive limit $\tau\to 0$, an effective thermal picture emerges and the $\hat{T}_\mathrm{bulk}$ can be expressed as,
\bea
\hat{T}_\text{bulk}=\frac{T_1^\mathrm{eff}+T_N^\mathrm{eff}}{2}\quad \mathrm{with}\quad T_i^\mathrm{eff}= \frac{v_0^2 \tau_i}{2\nu}. \label{effective_temp_clock}
\eea
Clearly, the effective temperature of the $i$-th reservoir, for the non-Markovian tumbling protocol, is reduced with respect to the Markovian case [see \eref{effective_temp}].

Finally, one can calculate the stationary state current $J_\mathrm{act}$ by substituting \eref{noise_clock_freq} in the Eq.~\eqref{average_energy_current} and integrating it numerically. In Fig.~\ref{fig_clock_current}(b) we have shown the analytical prediction of $J_\mathrm{act}$ with the numerical simulation. From Fig.~\ref{fig_clock_current}(b), it is clear that the most important qualitative features of the $J_\mathrm{act}$ namely the negative differential conductivity and the non-trivial sign reversal, mentioned in Sec.~\ref{stat_current_harmonic}, remain unaffected irrespective of change in the active force autocorrelation $\tilde h(\omega,\tau)$.

\subsection{Nonlinear active Rubin bath}

Finally, in this section, we explore the effect of non-linear interactions in the active reservoirs. In particular, we consider the interparticle potential $V(z)$ [see Eqs.~\eref{eomM} and \eref{eq:generic_prob}] to be of the famous Fermi-Pasta-Ulam-Tsingou form~\cite{fput_original,fput_dhar},
\bea
V(z)=\frac{\lambda}{2}z^2+\frac{\lambda_2}{4}z^4.\label{eq:fput_pot}
\eea 
\begin{figure}[ht]
\centering
    \includegraphics[width=\textwidth]{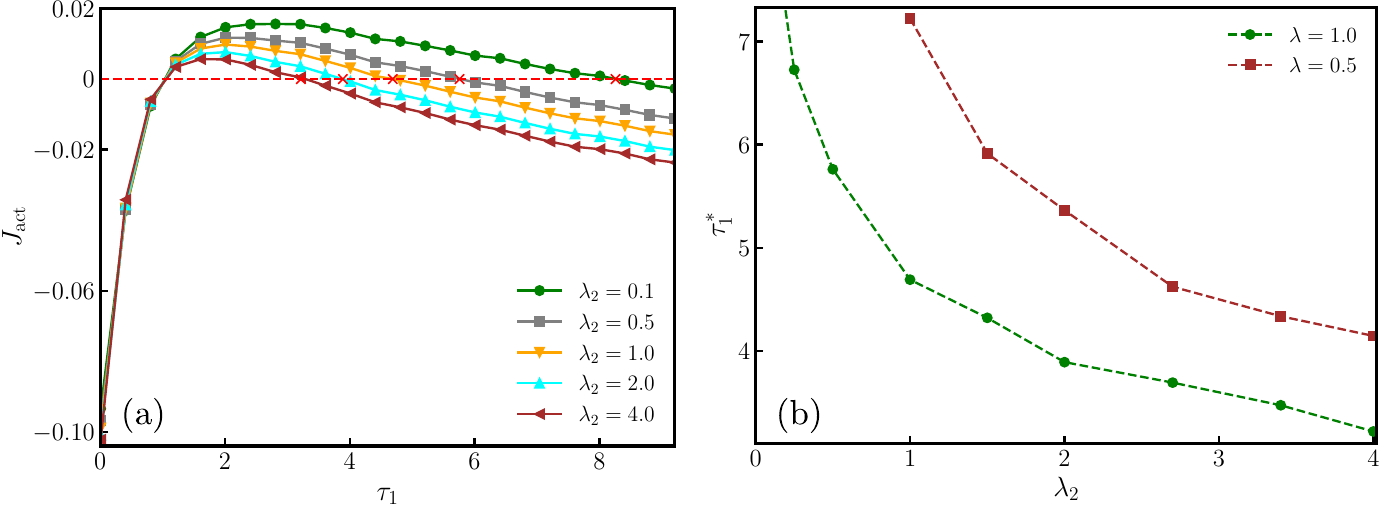} 
    \caption{Effect of non-linear interaction in the active reservoir: (a) Plot of $J_\mathrm{act}$ {\it vs} $\tau_1$, obtained from numerical simulations with different values of the non-linear coupling constant $\lambda_2$ for a fixed $\lambda=1.0$ and $\tau_N = 1$. 
    (b) Plot of the nontrivial current reversal point $\tau_1^*$, extracted from the numerical simulation data,  as a function of $\lambda_2$ for two different values of $\lambda$. The simulations are done with a system size $N=256$, bath size $M=256$ and $m=1=k=\nu=\lambda=v_0$.}
\label{fig:average_current_fput}
\end{figure}
Due to the non-linear nature of the corresponding equations of motion Eq.~\eqref{eomM}, it is difficult to obtain the effective equation of motion for the probe particle \eref{eq:generic_prob}. We use numerical simulations to measure the stationary current flowing through the system. Fig.~\ref{fig:average_current_fput}(a) shows the plot of $J_\text{act}$ for different values of the non-linear coupling constant $\lambda_2$. Clearly, the qualitative behavior of the current does not change---it shows a non-monotonic behavior as the activity drive is changed and undergoes a direction reversal at some non-trivial point $\tau_1^*$. However, this reversal point now depends on the non-linear coupling strength---$\tau_1^*$ decreases monotonically as $\lambda_2$ is increased. Figure~\ref{fig:average_current_fput}(b) shows a plot of $\tau_1^*(\lambda_2)$ for different values of $\lambda$.

\section{Conclusions}\label{sec:concl}
In this work, we propose a model for an active Rubin bath---a microscopic model for an active reservoir in the form of a harmonic chain of overdamped run-and-tumble particles. The activity of such a reservoir is characterized by the persistence time $\tau$ of the constituent particles, which are assumed to be the same. We characterize the behavior of this active reservoir by explicitly computing the dissipation and noise kernels experienced by a passive inertial probe connected to it. The active nature of the reservoir leads to a modification of the FDR which can not be described by an effective temperature picture in a thermodynamically consistent way.

We also study the properties of an ordered harmonic chain driven by two such active reservoirs with different activities $\tau_1$ and $\tau_N$. We characterize the NESS of the activity-driven system by computing the two-point correlation of the velocity of the bulk oscillators, kinetic temperature profile, and the average energy current flowing through the system. It turns out that the activity-driven NESS is characterized by several novel features compared to its thermally-driven counterpart. First, the active nature of the drive gives rise to a characteristic length scale $\ell\propto \max{(\tau_1,\tau_N)}$ over which the velocities of the bulk oscillators are correlated. This is in sharp contrast to the thermally-driven scenario where the velocity fluctuations of the bulk oscillators are uncorrelated. The two-time velocity correlation of a single oscillator also shows strong signatures of the activity in the short-time regime. Moreover, the average energy current shows a nonmonotonic behavior, accompanied by a nontrivial direction-reversal, as the activity drive is changed. It is to be noted that none of these behaviors, in general, can be explained by an effective temperature picture except in the passive limit $(\tau_1, \tau_N) \to 0$. We also perform numerical simulations with more generalized models for the active reservoirs by considering FPUT-type interactions among the reservoir particles and non-Markovian activity dynamics and find the same qualitative behavior of the energy current. 

The results obtained here and in some of our recent works \cite{activity_driven_chain,activity_stationary} suggest that the striking features of the current, namely, the non-monotonicity and the direction reversal are rather generic to the activity-driven harmonic systems. In this context, it would be interesting to investigate if other microscopic models of active reservoirs with hardcore or short-ranged interactions can lead to qualitative changes in the behavior of the energy current. Another relevant question is how the characteristic properties of the NESS change in the presence of disorder and correlated dynamics of the constituents of the reservoir particles. Finally, it is also worthwhile to investigate whether the qualitative behavior of the energy current changes in the presence of disorder and non-linearity in the driven system.

\acknowledgements{R. S. acknowledges support from the CSIR, India [Grant No. 09/0575(11358)/2021-EMR-I]. U.B. acknowledges support from the Science and Engineering Research Board (SERB), India, under a MATRICS grant [No. MTR/2023/000392].
}

\appendix
\section{Derivation of the generalized Langevin equation}\label{detail_active_chain}

In this section, we provide the details of the computation of the effective noise and the dissipation kernel acting on the passive probe particle [see Fig.~\ref{fig:single_particle}]. We start from \eref{eom_chain}, which can be conveniently recast in a matrix form,
\begin{align}
    \nu\dot{Y}(t)=\Psi Y(t)+W P(t)+F(t), \label{eq_act_dyn}
\end{align}
where $Y=\big(y_1(t), ~  y_2(t), \cdots y_{M}(t)\big)^T$ and $F=\big(f_1(t),~ f_2(t), \cdots f_M(t) \big)^T$. The information about the linear interaction of the active particles is encoded in the $M \times M$ tridiagonal matrix $\psi$ with elements
\bea
\Psi_{ij}=\lambda\left[\delta_{i+1,j}+\delta_{i,j+1}-2  \delta_{ij}\right]. 
\eea 
Finally, $P=\big(0,~0,~\cdots x_1(t)\big)^T$ encodes the position of the probe particle while matrix $W$ with the elements $W_{ij}=\lambda\delta_{Mj}$ denotes the coupling between the reservoir and the probe particle.

Our goal is to write an equation of motion for the probe particle, by integrating out the reservoir particle positions $\{ y_i(t) \}$. To this end, we first need to find the solution of \eref{eq_act_dyn} for a given $x_1(t)$, which can most conveniently be obtained by diagonalizing the tri-diagonal matrix $\Psi$ \cite{tridiagonal_matrix}. The eigenvalues of $\Psi$ are given by,
\bea
\mu_{k}=-4 \lambda \sin^2{\Big[ \frac{k \pi}{2(M+1)}\Big]},
\quad\text{where}\quad k=1,2\cdots M,
\eea
and the $j$-th component of the normalized eigenvector corresponding to $\mu_k$ is,
\bea
u_{j}^{(k)} = \sqrt{\frac 2{M+1}} \, \sin \frac{jk \pi}{M+1}. \label{matrix_U}
\eea
Thus, $\Psi$ is diagonalized by the similarity transformation,
\bea
D=U \Psi U^{-1},\label{similarity}
\eea 
where the diagonal matrix $D$ has the elements $D_{jk}=\mu_k \delta_{jk}$ and the diagonalizing matrix $U$ with elements $U_{jk} =u_{j}^{(k)}$ satisfies $U^2 = \mathbb{1}$. 

Multiplying \eref{eq_act_dyn} with $U$ from the left, we get,
\bea
\nu U \dot{{Y}}(t)= D UY(t) + U {W} P(t)+ U F(t),  \label{eq_ysolve}
\eea
which can be readily integrated to obtain,
\bea
{Y}(t) = \frac{1}{\nu}\int_{-\infty}^t ds   \Big[ U^{-1} e^{D\frac{(t-s)}{\nu}} U W P(s)+U^{-1} e^{D\frac{(t-s)}{\nu}}U F(s)\Big].\label{eom_matrix_active}
\eea
Using the explicit form of $W_{jk}$, $P_j(t)$ and $F_j(t)$, we finally get,
\bea
y_i(t)=\frac{ \lambda}{\nu} \int_{-\infty}^t ds \, x_1(s) \Lambda_{i  M}(t-s)  + \frac{1}{\nu}\int_{-\infty}^t ds \sum_{j=1}^{M} \Lambda_{i j}(t-s) f_{j}(s),
\eea 
where we have defined,
\bea 
\Lambda_{i j}(t)=\left(U^{-1}  e^{D t/\nu} U\right)_{ij}=\frac 2{M+1} \sum_{k=1}^{M} \sin \frac{i k \pi}{M+1} \sin \frac{j k \pi}{M+1}\exp{\Bigg[\frac{ \mu_k}{\nu}t\Bigg]},\label{eq_noise_matrix}
\eea 
which is also quoted in \eref{eq:lambda_jl} in the main text. Taking $i=M$ we get the equation of motion of $y_{M}(t)$, mentioned in \eref{bound_active}.

\section{Computation of the velocity correlation}\label{stationary_correl_app}

In this appendix, we provide the details of the computation for the two-point correlation of the velocities of the bulk oscillators. Using \eref{eq_vel_vel_correl}, the two-point correlation can be written as a sum of the contributions coming from the reservoirs as,
\bea
\la v_l(t) v_{l'}(t') \ra=\sum_{i=1,N} \chi_i(\tau_i,l,l',t,t'),
\eea 
with,
\begin{align}
\chi_i(\tau,l,l',t,t') \equiv \int_{-\infty}^{\infty} \frac{d\omega}{2 \pi} \omega^2 e^{-i \omega (t-t')} G_{li}(\omega)G_{l'i}^*(\omega)\tilde{g}(\omega,\tau).
\end{align}
Here $G(\omega)$ is the Greens's function matrix defined in \eref{ap_tridiag}. Because of the tridiagonal nature of $G^{-1}(\omega)$,  the elements of $G(\omega)$ can be computed explicitly~\cite{usmani}. The relevant elements required for our purpose are,
\bea
G_{l1}=(-k)^{l-1}\frac{\theta_{N-l}}{\theta_N}~~\text{and}~~
G_{lN}=(-k)^{N-l}\frac{\theta_{l-1}}{\theta_N}.\label{eq_G_matrix}
\eea
The explicit forms of $\theta_l$ for $l=0, 1, \dots N$ are given by~\cite{usmani,oscilatory_integral}, 
\begin{align}
\theta_0 &=1,\cr
\theta_1 &= -m\omega^2 + k-i\omega\tilde{\gamma}, \cr
\theta_l &= k^l\Big[ \cos{(lq)}+\frac{c(\omega)}{2 k \sin{q}} \sin{(lq)} \Big],\quad \forall l = 2,3,\cdot \cdot N-1
\quad \text{and} \cr
\theta_N &= k^{N-1}\Big[c(\omega) \cos{(Nq-q)}+d(\omega) \sin{(Nq-q)} \Big],\label{eq_theta_l_eq_theta_N}
\end{align}
where $\omega$ and $q$ are related by 
\bea
\omega=\omega_c \sin{\left(\frac{q}{2}\right)},\quad \text{with}\quad \omega_c = 2 \sqrt{ \frac{k}{m}}. \label{eq:wc_def}
\eea 
For notational simplicity, we have also defined, $c(\omega)=c_1(\omega)+i c_2(\omega)$ and $d(\omega)=d_1(\omega)+i d_2(\omega)$ in \eref{eq_theta_l_eq_theta_N} with,
\begin{align}
&c_1(\omega)=2\omega \mathrm{Im}[\tilde{\gamma}]  -m\omega^2,
\quad 
c_2(\omega)=-2\omega \mathrm{Re}[\tilde{\gamma}],\cr
&d_1(\omega)=\frac{\omega^2}{k \sin{q}}\left(\mathrm{Im}[\tilde{\gamma}]^2-\mathrm{Re}[\tilde{\gamma}]^2- m k \cos{q}-m\omega\mathrm{Im}[\tilde{\gamma}] \right),~~
d_2(\omega)=\frac{\omega^2 \mathrm{Re}[\tilde{\gamma}]}{k \sin{q}} \left(m \omega -2 \mathrm{Im}[\tilde{\gamma}]\right).~~~~~~~~~\label{eq_ref_c_d}
\end{align}

Using Eqs.~\eqref{eq_G_matrix} and \eqref{eq_theta_l_eq_theta_N} we can write the contribution from the left active reservoir as,
\begin{align}
&\chi_1(\tau,l,l',t,t')
=\int_{-\infty}^{\infty} \frac{d\omega}{2\pi} \frac{\omega^2 e^{-i \omega (t-t')}}{4 k^2 \sin^2{q}} \Bigg[\frac{\left(|c(\omega)|^2+4 k^2 \sin^2{q}\right)\cos{(l-l')q}}{\left|c(\omega) \cos{(N-1)q}+d(\omega) \sin{(N-1)q} \right|^2}\cr
 &-\frac{\left(|c(\omega)|^2-4 k^2 \sin^2{q}+4 k c_1\sin{q}\right)\sin{(l+l'-2 N)q}+i4k c_2 \sin{q} \sin{(l-l')q} }{\left|c(\omega) \cos{(N-1)q}+d(\omega) \sin{(N-1)q} \right|^2}\Bigg]\tilde{g}(\omega,\tau).\quad\label{ap_inegranf_vanish}
\end{align}
From \eref{eq:wc_def}, it is clear that, in the region $\omega>\omega_c$, $q$ becomes imaginary and  the integrand in Eq.~\eqref{ap_inegranf_vanish} vanishes exponentially as $e^{-2N\bar{q}}$  for real $\bar{q}=\pi -i q$. Therefore, for large $N$, non-zero contribution to the integral comes only from the region $|\omega|\leq\omega_c$, or $ |q|\leq \pi$. It is important to note that, the imaginary term present in \eref{ap_inegranf_vanish} vanishes as it turns out to be an odd function of $\omega$ (or $q$). Moreover, we are interested in the velocity correlation in the bulk, and without any loss of generality, we can take $l=\frac{N}{2}$, $l'=l+ \Delta l $. Thus, for large $N$ \eref{ap_inegranf_vanish} reduces to,
\begin{align}
\chi_1(\tau,\Delta l,t,t')
=\int_{-\omega_c}^{\omega_c} \frac{d\omega}{2\pi} \frac{\omega^2 e^{-i \omega (t-t')}}{4 k^2 \sin^2{q}} \Bigg(\frac{\left(|c(\omega)|^2+4 k^2 \sin^2{q}\right)\cos{(\Delta l q)}}{\left|c(\omega) \cos{(N-1)q}+d(\omega) \sin{(N-1)q} \right|^2}\cr
 -\frac{\left(|c(\omega)|^2-4 k^2 \sin^2{q}+4 k c_1\sin{q}\right)\sin{\big((N-1+\Delta l)q\big)} }{\left|c(\omega) \cos{(N-1)q}+d(\omega) \sin{(N-1)q} \right|^2}\Bigg)\tilde{g}(\omega,\tau),
\end{align}
which is a function of $\Delta l$ only. In the limit $N\to\infty$, one can average over the fast oscillations in $x=Nq$~\cite{oscilatory_integral} to get,
\begin{align}
\chi_1(\tau,\Delta l,t,t')=\int_{-\omega_c}^{\omega_c} \frac{d\omega}{2\pi}  \omega^2  e^{-i \omega (t-t')}\frac{\left(|c(\omega)|^2+4 k^2 \sin^2{q}\right)}{4 k^2 \sin^2{q}\Big(c_2 d_1-c_1 d_2\Big)}
\cos{(\Delta l q)}~~\tilde{g}(\omega,\tau).\label{eq:even_integral_spatial}
\end{align}
Note that, here we have used the identities,
\begin{align}
&\int_0^{2\pi} \frac{dx}{2 \pi} \frac{1}{t_1 \sin^2x+t_2 \cos^2x+t_3 \sin{x}\cos{x}}
=\frac{2}{\sqrt{4 t_1 t_2-t_3^2}},\cr
&\int_0^{2\pi} \frac{dx}{2 \pi} \frac{\cos{x}}{t_1 \sin^2x+t_2 \cos^2x+t_3 \sin{x}\cos{x}}=
\int_0^{2\pi} \frac{dx}{2 \pi} \frac{\sin{x}}{t_1 \sin^2x+t_2 \cos^2x+t_3 \sin{x}\cos{x}}=0.~~~~~~~~~\label{eq_oscillatoryintegral}
\end{align}
Using the explicit expression of $c_1,c_2,d_1,d_2$ given in Eq.~\eqref{eq_ref_c_d} we get, 
\begin{align}
c_2 d_1-c_1 d_2
= \frac{2\omega^3\,\mathrm{Re}[\tilde{\gamma}] (m k +|\tilde{\gamma}|^2 - m\omega\,\mathrm{Im}[\tilde{\gamma}])}{k \sin{q}}. \label{eq_g1_2}
\end{align}
The contribution from the right active reservoir can also be calculated in a similar manner. Finally, we arrive at \eref{eq:general_vel_correl} where we have used the fact that $\mathrm{Re}[\tilde \gamma(-\omega)]=\mathrm{Re}[\tilde \gamma(\omega)]$ and $\mathrm{Im}[\tilde \gamma(-\omega)]=-\mathrm{Im}[\tilde \gamma(\omega)]$
 and the explicit form of $\tilde{g}(\omega,\tau)$ given in \eref{spectra_define}.

\section{Computation of the average stationary current} \label{appendix_formalism}

The average energy current in the steady state can be written as,
\bea 
J_\text{act}=J_1+J_2, \label{total_current}
\eea 
where, 
\bea
J_1 = \Big\la \Big(-\int_{-\infty}^{t} ds \, \dot{x}_1(s)\gamma(t-s)\Big) \dot{x}_1(t)\Big\ra
\quad \mathrm{and,}\quad
J_2 = \Big\la \Sigma_1(t) \dot{x}_1(t)\Big\ra. 
\eea
It is convenient to recast these quantities in a matrix notation,
\bea
J_1 = -\Big\la \mathrm{Tr}\Big[ \dot{X}(t)\int^t_{-\infty}ds \dot{X}^T(s)\Gamma_1(t-s)\Big] \Big\ra \quad \mathrm{and,}\quad
J_2 = \Big\la \mathrm{Tr}\Big[\Xi_1(t) \dot{X}^T(t) \Big] \Big\ra, \label{eq_j2_matrix}
\eea
where we have defined,
\begin{align}
&[\Gamma_{1}(t)]_{ij}=\gamma(t) \delta_{i 1} \delta_{j 1},\quad [\Gamma_{N}(t)]_{ij} =\gamma(t) \delta_{i N}\delta_{j N},\label{eq:Gam_def}\\
&[\Xi_{1}(t)]_{j} = \Sigma_1(t) \delta_{1j},\quad [\Xi_{N}(t)]_{j}=\Sigma_N(t) \delta_{Nj},\quad\text{and}\\
&\Xi(t)=\Xi_1(t)+\Xi_N(t). \label{part_def}
\end{align}
In the following, we evaluate $J_1$ and $J_2$ separately. Using Eq.~\eqref{fourier_sol} in \eref{eq_j2_matrix}, we have,
\bea
J_1 
=\int \frac{d\omega }{2 \pi}\frac{ d \omega'}{2 \pi} \omega \omega' e^{-i(\omega+\omega')t}\mathrm{Tr}\Big[G(\omega) \Big\la \tilde{\Xi}(\omega) \tilde{\Xi}(\omega') \Big\ra\tilde{\Gamma}_1(\omega') \Big],\label{j1_1}
\eea
where $\tilde \Xi(\omega)$ is the Fourier transform of $\Xi(t)$ defined in \eref{part_def}. Next, from \eref{correlation_freq}, we have,
\bea
\la \tilde{\Xi}(\omega) \tilde{\Xi}(\omega') \ra = 2\pi \delta(\omega+\omega')\left(S_1(\omega)+S_N(\omega)\right) ~~ \mathrm{where,}~~~
[S_{i}(\omega)]_{kl}=\delta_{ik} \delta_{il} \, \tilde g(\omega, \tau_i).
\label{eq_SL_SR}
\eea
Substitution of \eqref{eq_SL_SR} in \eref{j1_1} yields,
\begin{align}
J_1=-\int^{\infty}_{-\infty} \frac{d \omega}{2 \pi} \omega^2 \Big[G(\omega)\Big(S_1(\omega)+S_N(\omega)\Big)\tilde{\Gamma}_1(-\omega) \Big]. \label{eq_j1}
\end{align}
We can proceed similarly to evaluate $J_2$, which leads to,
\bea
J_2=\frac{i}{2\pi}\int^{\infty}_{-\infty} d\omega \, \omega \, \mathrm{Tr}\Big[G(-\omega)S_1(\omega) \Big]. \label{eq_j2}
\eea
Thus, using Eqs.~\eqref{eq_j1} and \eqref{eq_j2} in \eref{total_current}, the total average energy current in the stationary state can be expressed as,
\begin{align}
J_\text{act}=\int^{\infty}_{-\infty} \frac{d \omega}{2 \pi} \omega \mathrm{Tr}\Big[\Big(i G^* -\omega G^{*} \tilde{\Gamma}_1^{*} G\Big) S_1 \Big]
- \int_{-\infty}^{\infty} \frac{d\omega}{2 \pi} \omega^2 \mathrm{Tr}\Big[ G^{*} \tilde{\Gamma}^{*}_1G S_N \Big],\label{eq_J_act}
\end{align}
where we have separated the terms containing $S_1$ and $S_N$. As shown below, the above equation can be further simplified by exploiting properties of $G(\omega)$. From Eq.~\eqref{ap_tridiag}, we have,
\bea
G^{-1} = -\omega^2 M-i\omega(\tilde{\Gamma}_1+\tilde{\Gamma}_N)+\Phi. \label{eq_invG}
\eea
Taking the complex conjugate of $G^{-1}$ and subtracting it from Eq.~\eqref{eq_invG}, we arrive at,
\bea
\omega G^* \tilde{\Gamma}^{*}_1G = -i (G-G^*)-\omega G^*(\tilde{\Gamma}_N+\tilde{\Gamma}^{*}_N+\tilde{\Gamma}_1)G. \label{eq_1}
\eea
Using Eq.~\eqref{eq_1} in Eq.~\eqref{eq_J_act}, we can rewrite the term containing $S_1$ as,
\begin{align}
H_1 = \int^{\infty}_{-\infty} \frac{d\omega}{2 \pi}\omega \mathrm{Tr}[i(G+G^*)S_1]
+\int^{\infty}_{-\infty} \frac{d\omega}{2 \pi} \mathrm{Tr}\left[G^*\left(-i\omega \mathbb{1} +\omega^2(\tilde{\Gamma}_1+\tilde{\Gamma}_N+\tilde{\Gamma}^{*}_N)\right)G S_1\right]. \label{h1_1}
\end{align}
The first term vanishes since the integrand is an odd function of $\omega$. 
Moreover, multiplying \eref{eq_invG} with $i \omega G$ from right, we get,
\bea
-i\omega \mathbb{1}+\omega^2 (\tilde{\Gamma}_1+\tilde{\Gamma}_N)G=i \omega^3 M G-i\omega \Phi G.\label{simplification}
\eea
Substituting \eref{simplification} in \eref{h1_1} we arrive at,  
\begin{align}
H_1 =\int^{\infty}_{-\infty} \frac{d\omega}{2 \pi} \mathrm{Tr}\left[G^*\left(i\omega^3 M -i \omega \Phi \right)G S_1\right]+\int^{\infty}_{-\infty} \frac{d\omega}{2 \pi} \omega^2 \mathrm{Tr}\left[G^*\tilde{\Gamma}^{*}_NG S_1\right].
\end{align}
Once again, the first integral in the above equation vanishes  
as the integrand is an odd function of $\omega$. Therefore, we are left with only one term that contains $S_1$. Using \eref{eq_J_act} and combining the contributions from both reservoirs, we arrive at,
\bea
J_\text{act}=\int_{-\infty}^{\infty} \frac{d \omega}{2 \pi} \omega^2
 \mathrm{Tr}\Big[G^* \tilde{\Gamma}^{*}_N G S_1-G^* \tilde{\Gamma}_1 G S_N \Big].
\eea
Using explicit forms of $\tilde \Gamma_i(\omega)$ and $S_i(\omega)$ from Eqs.~\eqref{eq:Gam_def} and \eqref{eq_SL_SR}, and taking the trace, we arrive at a `Landauer-like' formula,
\begin{align}
J_\text{act}=\int_{-\infty}^{\infty} \frac{d \omega}{2 \pi} \omega^2 |G_{1N}|^2 \mathrm{Re}[\tilde{\gamma}(\omega)]\Big(\tilde{g}(\omega,\tau_1)-\tilde{g}(\omega,\tau_N)\Big).\label{jact_app}
\end{align}
The matrix element $G_{1N} = \theta_N^{-1}$ [see \eref{eq_G_matrix}], where $\theta_N(\omega)$ is given by   \eref{eq_theta_l_eq_theta_N}. For thermodynamically large system size $N \to \infty$, the integrand in \eref{jact_app} is non-zero only within the band $|\omega|<\omega_c$ [see the discussion after Eq.~\eqref{ap_inegranf_vanish}]. Furthermore, in this thermodynamic limit, one can also integrate out the fast oscillations using Eq.~\eqref{eq_oscillatoryintegral}, which leads to a rather simple expression,
\bea
|G_{1N}|^2=(c_2 d_1-c_1 d_2)^{-1}\label{eq_g1_1}=\frac{k \sin{q}}{2\omega^3\,\mathrm{Re}[\tilde{\gamma}] (m k +|\tilde{\gamma}|^2 - m\omega\,\mathrm{Im}[\tilde{\gamma}])}.\label{g1N}
\eea
Note that, in the last step we have used \eref{eq_g1_2}. Substituting \eref{g1N} and \eref{spectra_define} in 
\eref{jact_app} we arrive at \eref{average_energy_current} in the main text.

\bibliography{bibfile}

\end{document}